\documentclass[12pt,a4paper]{article}
\usepackage{amsmath,amssymb,dsfont,epsfig,a4wide}
\usepackage{lscape,longtable,xcolor}
\usepackage{extarrows}
\usepackage{booktabs,multirow,makecell}
\usepackage{float}
\usepackage[round]{natbib}
\usepackage{hyperref}
\usepackage{graphicx}
\usepackage{caption,subcaption}

\usepackage{ulem}
\allowdisplaybreaks

\newcommand{\var}{\mbox{var}}
\renewcommand{\hat}{\widehat}

\newtheorem{Th}{{\bf Theorem}}

\newtheorem{asu}{Condition }
\newcounter{subassumption}[asu]
\renewcommand{\thesubassumption}{(\textit{\roman{subassumption}})}
\makeatletter
\renewcommand{\p@subassumption}{\theasu}
\newcommand{\subasu}{
	\refstepcounter{subassumption}
	\thesubassumption}

\newtheorem{corollary}{Corollary}
\newtheorem{lemma}{Lemma}

\newcommand{\calW}{{\cal W}}

\def\lmax{\lambda_{\tiny\max}}

\def\bse{\begin{eqnarray*}}
	\def\ese{\end{eqnarray*}}
\def\be{\begin{eqnarray}}
\def\ee{\end{eqnarray}}
\def\bsq{\begin{equation*}}
\def\esq{\end{equation*}}
\def\bq{\begin{equation}}
\def\eq{\end{equation}}

\def\E{\hbox{E}}
\def\var{\hbox{var}}

\def\wh{\widehat}
\def\wt{\widetilde}
\def\th{^{\mbox{\tiny\rm th}}}

\def\AIC{\mbox{AIC}}
\def\AICC{\mbox{AICC}}
\def\SAICC{\mbox{SAICC}}
\def\BIC{\mbox{BIC}}

\def\NMSPE{\mbox{NMSPE}}

\def\argmin{\mbox{argmin}}

\def\tr{\mbox{trace}}

\def\bb{{\boldsymbol\beta}}
\def\bbeta{{\boldsymbol\beta}}

\def\0{{\bf 0}}
\def\1{{\bf 1}}
\def\A{{\bf A}}
\def\H{{\bf H}}

\def\V{{\bf V}}

\def\K{{\bf K}}

\def\w{{\bf w}}

\def\X{{\bf X}}
\def\x{{\bf x}}

\def\y{{\bf y}}
\def\e{{\bf e}}

\def\bmu{{\boldsymbol \mu}}
\def\wh{\widehat}
\def\wt{\widetilde}
\def\var{\hbox{var}}

\def\sums{\sum\nolimits_{s=1}^{{S_n}}}

\def\s{{(s)}}

\def\tt{^{\rm T}}

\newcommand\sumjn{\sum_{j=1}^{n}}
\newcommand\sumin{\sum_{i=1}^{n}}
\newcommand\sumjsn{\sum_{j^\ast=1}^{n}}
\newcommand\sumsz{\sum_{s=1}^{S_0}}
\newcommand{\infs}{\min_{1\leq s\leq S_n}}
\newcommand{\infsj}{\infs\min_{1\leq j \leq J_n}}

\newcommand\kernel{k_{h_s}}

\newcommand\js{{j^\ast}}
\newcommand{\hbeta}{\hat{\boldsymbol\beta}}
\newcommand{\hhbeta}{{\hat{\boldsymbol\beta}^{R}}}
\newcommand{\hhbetaj}{{\hat{\boldsymbol\beta}^{R [-j]}}}
\newcommand{\sbeta}{\boldsymbol{\beta}^\ast}
\newcommand{\shbeta}{\boldsymbol{\beta}^{R\ast}}
\newcommand{\onen}{\frac{1}{n}}
\newcommand\setA{\mathcal{A}}
\newcommand\setB{\mathcal{B}}
\newcommand{\sumsn}{\sum_{s=1}^{S_n}}
\newcommand{\sups}{\max_{1\leq s\leq S_n}}
\newcommand{\supi}{\max_{1\leq i\leq n}}
\newcommand{\supj}{\max_{1\leq j\leq n}}
\newcommand{\supij}{\max_{1\leq i\leq n}\max_{1\leq j\leq n}}
\newcommand{\supjn}{\max_{1\leq j\leq J_n}}

\newcommand{\bSigma}{{\bf \Sigma}}
\newcommand{\md}{\,\mathrm{d}}

\renewcommand{\theequation}{\arabic{section}.\arabic{equation}}
\renewcommand{\thesubsection}{\arabic{section}.\arabic{subsection}}

\def\boxit#1{\vbox{\hrule\hbox{\vrule\kern6pt
			\vbox{\kern6pt#1\kern6pt}\kern6pt\vrule}\hrule}}

\begin{document}
	\title{\textbf{Optimal model averaging for single-index models with divergent dimensions}}
	\author{Jiahui Zou\footnote{School of Statistics, Capital University of Economics and Business},\quad Wendun Wang\footnote{Econometric Institute, Erasmus University Rotterdam; Tinbergen Institute}, \quad Xinyu Zhang\footnote{Academy of Mathematics and Systems Science, Chinese Academy of Sciences}\quad
		and Guohua Zou\footnote{School of Mathematical Sciences, Capital Normal University}}
	
	\date{March 1, 2020}
	\maketitle
	
	\begin{abstract}
		This paper offers a new approach to address the model uncertainty in (potentially) divergent-dimensional single-index models (SIMs). We propose a model-averaging estimator based on cross-validation, which allows the dimension of covariates and the number of candidate models to increase with the sample size. We show that when all candidate models are misspecified, our model-averaging estimator is asymptotically optimal in the sense that its squared loss is asymptotically identical to that of the infeasible best possible averaging estimator. In a different situation where correct models are available in the model set, the proposed weighting scheme assigns all weights to the correct models in the asymptotic sense.
		We also extend our method to average regularized estimators and propose pre-screening methods to deal with cases with high-dimensional covariates.
		We illustrate the merits of our method via simulations and two empirical applications.

	\end{abstract}
	
	
	
	\clearpage
	\section{Introduction}\label{chap:introduction}
	A linear regression model is a common tool to analyze the relationship between a response variable of interest $y$ and a vector of covariates $\x$ in diversified fields. However, in many applications, such a relationship is nonlinear, e.g., \cite{naik:tsai:2001} shows that car pricing nonlinearly depends on the number of car features; \cite{liang.wang.carroll:2007} documents a nonlinear relationship between the HIV viral load and the treatment time in a study of the effectiveness of antiretroviral medicines.
	%
	A natural extension to relax linearity is to consider a single-index model (SIM) that enables $y$ to depend on $\underline{\x}$ via an unknown and possibly nonlinear link function $g$, i.e.,
	\begin{align}\label{eq:model}
	y=g(\underline{\x}\tt \underline{\bbeta})+\underline{\epsilon},
	\end{align}
	where $\underline{\bbeta}$ is a 
	vector of unknown parameters, and $\underline{\epsilon}$ is the disturbance term. With an unknown link function, this model avoids the curse of dimensionality in many nonparametric models and reduces the risk of model misspecification, while maintaining relative ease of interpretation {
		\citep{Horowitz1998Semiparametric,naik:tsai:2001}}.
	Various approaches have been proposed to estimate the SIM, e.g., average derivative estimation \citep{PowellSemiparametric}, nonlinear least squares \citep{Ichimura1993Semiparametric}, and profile least-squares \citep{LiangEstimation}. All of these methods require correct specification of the covariates.
	However, this knowledge is often unavailable in practice, especially when there are many covariates, so researchers are exposed to a potentially large degree of model uncertainty with respect to which covariates should be included in the model.

	
	A popular method to address model uncertainty is model selection, which  picks the ``best'' model based on certain data-driven criteria, e.g., information criteria. Traditional model selection methods have been extended to SIMs, such as AIC \citep{naik:tsai:2001} and cross-validation \citep{kong.xia:2007}. More recently, \citet{cheng.zeng:2017} studied a shrinkage-type estimator to select and estimate covariates in SIMs.
	%
	{
		As an alternative to model selection, model averaging (MA) addresses model uncertainty by combining estimators from all candidate models with certain weights based on the model performance, and it often leads to lower risk than model selection \citep{hansen:2014}.}
	The past decade has witnessed burgeoning literature pertaining to model averaging. There are two main streams of averaging techniques: Bayesian model averaging (BMA) and frequentist model averaging (FMA). Although BMA is flexible and can be applied in many models, the choice of prior probabilities is often challenging and experiential \citep[see][for an excellent overview]{hoeting.madigan.ea:1999}. There are various FMA methods, and a partial list includes smoothed information criteria \citep[e.g.,][]{Buckland:1997}, optimal averaging \citep[e.g.,][]{hansen2007least}, adaptive combination \citep[e.g.,][]{yuan.yang:2005} and plug-in methods \citep[e.g.,][]{LiuDistribution}. Despite the increasing popularity of the averaging techniques, MA estimators for SIMs are hardly studied.

	This paper proposes a new model-averaging estimator to address model uncertainty in SIMs. We focus on prediction of the response variable and combine the predictions from multiple models with certain weights. Thus, we employ the optimal averaging method, which aims to achieve an averaged prediction that outperforms any single-model prediction.
	The proposed SIM averaging offers a flexible method to predict the response variable and explicitly considers the model uncertainty.
	To appropriately choose the averaging weights, we adopt a cross-validation criterion. One of the advantages of this criterion is that it does not require an unbiased estimator of risk, which is difficult to obtain for SIMs and is easy to implement.
	We justify the proposed estimator in two cases. First, we show that when all candidate models are misspecified, our resulting weight vector is asymptotically optimal to achieve the minimum squared loss as the infeasible best possible model-averaging estimator. Thus, our method can produce better prediction than other averaging methods for large samples even if the true data-generating process (DGP) is not an SIM, or certain covariates are not available to researchers.
	Second, when the set of candidate models includes correct models which nest the DGP (see Section~\ref{sec:Asymptotic} for more precise definition), we show that our weight choice can consistently ``pick'' the correct models in the sense that the sum of the weights assigned to the correct models tends to one when the sample size increases. If the bias caused by misspecification (omitting variables) is not asymptotically diminishing, the consistent weight choice further implies better prediction than any other averaging estimates based on misspecified models.
	An important merit of our approach is that in both cases, we allow the dimension of covariates and the number of candidate models to diverge as the sample size increases. { Moreover, we also study how to perform model averaging in high-dimensional situations, where the number of covariates is overly large and may even exceed the number of observations, such that estimating and averaging over all candidate models are infeasible. We propose averaging regularized estimators with an $L_1$ penalty as well as model screening methods. We establish the asymptotic properties of the averaging estimator based on regularization.}

	We contribute to the model averaging literature in three main respects. First, we propose a new model-averaging estimator for single-index models and establish its asymptotic optimality in terms of minimum squared loss.
	{Despite a wide range of FMA applications in different models, relatively fewer studies have considered semi- or nonparametric optimal model averaging.} \citet{li.li.racine.zhang:2018} considered  averaging varying-coefficient models. \citet{liu:2018} considered (non-optimal) model averaging for kernel regressions. \citet{zhang.wang:2019} studied optimal model averaging in partially linear models. \citet{ZhuA} further studied optimal averaging in varying-coefficient partially linear models. {\citet{Liu2021} studied model averaging of varying-coefficient models, where the varying coefficients are modelled as a single-index function.} 
	To the best of our knowledge, this paper is the first {comprehensive study on the properties} of optimal model averaging for single-index models. {Compared with the existing semi-/non-parametric model averaging literature, we allow the number of candidate models and the dimension of each candidate model to diverge when the sample size increases, which is particularly useful in a high-dimensional setting.}  Our study also complements \citet{hansen2014book} by formally establishing theoretical properties of the averaging estimator for nonparametric models based on a cross-validation criterion.

	Second, {we study} the asymptotic properties of the SIM averaging estimator when the candidate models include correct models. In the framework of linear models, \citet{zhang2019Parsimonious} showed the consistency of averaging coefficient estimators when there is at least one correct model in the candidate model set. However, no asymptotic results have been established for semi- or nonparametric model averaging when correct models are available.
	We fill in this gap by providing the asymptotic behavior of our weight estimators for SIMs. We show that our averaging method can consistently choose the correct models by asymptotically assigning all weights to the correct models, {no matter whether the candidate models are of finite or diverging dimension}. This result complements the asymptotic optimality when all candidate models are misspecified and demonstrates the validity of our method when there are correct models in the model space.

	{Last but not least, this paper deals with high-dimensional model averaging, and offers the first study on the properties of averaging \emph{regularized} estimators. While \citet{zhang2019Parsimonious} advocated the use of regularized estimators for preparing candidate models in linear model averaging with many covariates, no theoretical properties of such averaging estimators are provided and it is not clear whether the regularization-based averaging is still asymptotically optimal. Our analysis leads to an affirmative answer, thus providing justifications for averaging  regularized estimators. This approach allows us to deal with the cases in which there are more parameters to estimate than the available observations. Based on the regularized estimation, we also propose a preliminary model screening procedure to shrink the candidate model space. Our methods to deal with high-dimensionality differ from that of \citet{Ando2014}, which reduced the dimension by grouping the covariates and only averaging estimators associated with pre-selected  groups.}

	Our theories are verified via an extensive set of simulation experiments. We also apply the proposed method to a real dataset, which revisits the relationship between financial development and income distribution using cross-country data. 

	The remainder of this paper is organized as follows. Section~\ref{sec:MA} introduces the model-averaging method for SIMs. Section~\ref{sec:asp_prop} studies its theoretical properties.
	Section~\ref{sec:pbigern} considers averaging based on regularized estimation and preliminary model screening { to deal with a large number of covariates.}
	Section~\ref{sec:simulation} presents the simulation study, and Section~\ref{sec:empirical} provides the empirical application. Section~\ref{sec:conclusion} concludes with some remarks. Appendix provides additional conditions needed for lemmas, theorems and corollaries. The Online Supplement contains more detailed discussions of theoretical results, proofs, related methods, additional simulation studies, and another empirical application.

	\section{Model setup and estimation}
	\label{sec:MA}
	
	This section first sets up the model, and then provides a new averaging method to address model uncertainty and promote prediction ability for SIMs.
	
	\subsection{Single-index model averaging}
	\label{sec:model}
	Assume the following DGP, which is also referred to as the true model:
	\begin{align}
	y_i=\mu_i+\epsilon_i,\qquad i=1,2\ldots,n\nonumber,
	\end{align}
	where $y_i$ is the response variable of interest with mean $\mu_i$, and the random disturbances $\epsilon_1,\epsilon_2,\ldots,\epsilon_n$
	are independent and (possibly) heteroscedastic with
	$\E(\epsilon_i)=0$ and $\E(\epsilon_i^2)=\sigma_i^2$.
	Our purpose is to estimate $\mu_i$ and {thus predict the response variable} with $p$-dimensional covariates $\x_i=(x_{1,i},x_{2,i},\ldots,x_{p,i})\tt$ for $i=1,2,\ldots,n$, {where $p$ is allowed to be finite or divergent when the sample size $n$ increases.} {Besides, $\x_i$ is independent with $\epsilon_j$ for any $i,j=1,2,...,n$}.
	To this end, one may employ an SIM that enables us to flexibly model the dependence of $\mu_i$ on $\x_i$. However, it is unclear in practice which covariates in $\x_i$ should be used for the prediction. Including unnecessary covariates may cause efficiency loss and impose heavier computational burden, especially in the nonparametric context.

	We propose to tackle the model uncertainty by averaging the estimators obtained from various candidate models, each of which includes a distinct subset of covariates.
	In particular, suppose that we have $S_n$ candidate models in total with distinct specifications of covariates. The $s\th$ candidate model can be written as
	\begin{align}
	y_i&= g_\s(\x_{\s,i}\tt\bbeta_\s) + \epsilon_{\s,i},\quad
	i=1,2,\ldots,n,\quad s=1,2,\ldots,S_n,\notag
	\end{align}
	where $\x_{\s,i}$ is the $p_s$-dimensional covariate vector whose elements are a subset of $\x_i$, $\bbeta_\s$ is the associated parameter vector, $g_{\s}(\cdot)$ is an unknown link function that is allowed to vary across models, and $\epsilon_{\s,i}=y_i-g_\s(\x_{\s,i}\tt\bbeta_\s)$. As $p$ is potentially divergent, $p_s$ may also diverge as $n\to\infty$ for some $s\in\{1,2,\ldots,S_n\}$, leading to an increasing dimension of the parameter vector $\bbeta_\s$ in the candidate models.
	%
	%
	
	To estimate each candidate SIM, we follow \citet{Ichimura1993Semiparametric} {to achieve the identification of $\bbeta_{(s)}$ by {normalizing its first element to 1} 
		and employ the nonlinear least squares (NLS).} One of the advantages of NLS is its light computation burden, which is crucial in our case since our averaging technique is based on a cross-validation criterion (discussed in detail below) and the number of candidate models is typically substantial.
	To define the NLS estimator, let $k(\cdot)$ be a kernel function. For the $s\th$ candidate model, denote $h_s$ as the bandwidth, $k_{h_s}(\cdot)=k(\cdot/h_s)/h_s$, and $\K_\s(\bb_\s)=\{K_{\s,ij}(\bb_\s)\}_{n\times n}$ as an $n\times n$ smoother matrix with the typical element
	\be
	K_{\s,ij}(\bb_\s)=k_{h_s}(\x_{\s,i}\tt\bb_\s -
	\x_{\s,j}\tt\bb_\s)/
	\sum\nolimits_{j^\ast=1}^nk_{h_s}(\x_{\s,i}\tt\bb_\s-
	\x_{\s,j^\ast}\tt\bb_\s).\nonumber\ee
	Further define $\y=(y_1,y_2,\ldots,y_n)\tt$, $\bmu=(\mu_1,\mu_2,\ldots,\mu_n)\tt$, and $\X_\s=(\x_{\s,1}, \x_{\s,2},\ldots,\x_{\s,n})\tt$. The NLS estimator $\wh\bb_\s$ for the $s\th$ candidate model can then be obtained by minimizing the following objective function:
	\begin{align}
	H_{\s,n}(\bbeta_\s)=n^{-1}\left\|\y-\K_\s(\bb_\s)\y\right\|^2.\label{eq:objective}
	\end{align}
	The resulting estimator of $\bmu$ from the $s\th$ candidate model is $\wh\bmu_\s = \K_\s(\wh\bb_\s)\y$.
	With the estimator of each candidate model, we can obtain the model
	averaging estimator of $\bmu$ as
	\begin{align}
	\wh\bmu(\w)=\sums w_s\wh\bmu_\s=\K(\w,\wh\bb)\y,
	\label{whmu}
	\end{align}
	where $\wh\bb=(\wh\bb_{(1)}\tt,\wh\bb_{(2)}\tt,\ldots,\wh\bb_{(S_n)}\tt)\tt$, $\K(\w,\wh\bb)=\sums w_s\K_\s(\wh\bb_\s)$, and the weight vector $\w=(w_1,w_2,\ldots,w_{S_n})\tt$  belongs to the set $\calW = \left\{ \w \in {{\left[ {0,1}
			\right]}^{S_n}}:\sum\nolimits_{s = 1}^{S_n}w_s  = 1 \right\}$.
	The averaging estimator $\wh\bmu(\w)$ offers an appealing method to predict the response variable.
	
	Despite our primary goal of prediction, we are also interested in the functional effect of covariates (jointly depicted by the estimated coefficients and link function) if any of the candidate models is correctly specified, i.e., the data are generated by a single-index model with covariates being a subset of $\x_i$.
	Hence, we aim at developing a weight choice method, which not only provides optimal prediction but also consistently selects the correctly specified models if they exist in the set of candidate models.

	\subsection{Choosing the averaging weights}
	Given our main goal of prediction, our weight choice aims at minimizing the squared loss $L_n(\w)=\left\|\wh\bmu(\w)-\bmu\right\|^2$. We propose to choose the averaging weights using $J$-fold cross-validation (CV) in a similar manner to the jackknife model average of \cite{hansen.racine:2012}. CV is a numerical model-averaging method that is easy to implement and hardly relies on the structure of the model, except the dependence features of data. Unlike the Mallows criterion, this method is more flexible, since it does not require an unbiased estimator of risk, which is often difficult to obtain for complex models such as the single-index models considered here.
	
	To implement the $J$-fold CV, we divide the dataset into $J_n$ blocks so that there are $M_n=\lfloor n/J_n\rfloor$ observations in each block, where $\lfloor\cdot\rfloor$ denotes the integer part of a number. 
	For the $s\th$ candidate model, let $\wh\bb_\s^{[-j]}$ be the NLS estimator of $\bb_\s$ without using the observations from the $j\th$ block for $j=1,2,\ldots,J_n$.
	Then, the corresponding leave-block-out kernel estimator is $\wt\bmu_\s=(\wt\mu_{\s,1}, \wt\mu_{\s,2},\ldots.,\wt\mu_{\s,n})\tt$ with
	$$
	\wt\mu_{\s,1}
	=
	\begin{pmatrix}
	\0_{M_n}\tt,
	\frac{k_{h_s}\left(\x_{\s,M_n+1}\tt\wh\bb_\s^{[-1]} -
		\x_{\s,1}\tt\wh\bb_\s^{[-1]}\right)}{\sum\limits_{M_n<i\leq n}k_{h_s}\left(\x_{\s,i}\tt\wh\bb_\s^{[-1]} -
		\x_{\s,1}\tt\wh\bb_\s^{[-1]}\right)},
	\cdots,
	\frac{k_{h_s}\left(\x_{\s,n}\tt\wh\bb_\s^{[-1]} -
		\x_{\s,1}\tt\wh\bb_\s^{[-1]}\right)}{\sum\limits_{M_n<i\leq n}k_{h_s}\left(\x_{\s,i}\tt\wh\bb_\s^{[-1]} -
		\x_{\s,1}\tt\wh\bb_\s^{[-1]}\right)}
	\end{pmatrix}
	\tt
	\y,
	$$
	$$
	\vdots
	$$
	$$
	\wt\mu_{\s,n}
	=
	\begin{pmatrix}
	\frac{k_{h_s}\left(\x_{\s,1}\tt\wh\bb_\s^{{[-J_n]}} -
		\x_{\s,n}\tt\wh\bb_\s^{{[-J_n]}}\right)}{\sum\limits_{1\leq i\leq n-M_n}k_{h_s}\left(\x_{\s,i}\tt\wh\bb_\s^{{[-J_n]}} -
		\x_{\s,n}\tt\wh\bb_\s^{{[-J_n]}}\right)},
	\cdots,
	\frac{k_{h_s}\left(\x_{\s,n-M_n}\tt\wh\bb_\s^{{[-J_n]}} -
		\x_{\s,n}\tt\wh\bb_\s^{{[-J_n]}}\right)}{\sum\limits_{1\leq i\leq n-M_n}k_{h_s}\left(\x_{\s,i}\tt\wh\bb_\s^{{[-J_n]}} -
		\x_{\s,n}\tt\wh\bb_\s^{{[-J_n]}}\right)},
	\0_{M_n}\tt
	\end{pmatrix}
	\tt
	\y,
	$$\\
	where $\0_{M_n}$ is an $M_n$-dimensional vector of zeros.
	The above equations suggest that there is a matrix $\wt\K_\s(\wt\bb_\s)$ with $\wt\bb_\s=({\wh\bb_\s^{[-1]}}{\tt},{\wh\bb_\s^{[-2]}}{\tt},\ldots,{\wh\bb_\s^{[-J_n]}}{\tt})\tt$, such that
	the leave-block-out estimator of $\bmu$ under the $s\th$ candidate model can be written as $\wt\bmu_\s = \wt\K_\s(\wt\bb_\s)\y$.
	Let $\wt\bb=(\wt\bb_{(1)}\tt,\wt\bb_{(2)}\tt,\ldots,\wt\bb_{(S_n)}\tt)\tt$ and  $\wt\K(\w,\wt\bb)=\sums w_s\wt\K_\s(\wt\bb_\s)$.
	The averaging leave-block-out estimator of $\bmu$ is then given by
	\be
	\wt\bmu(\w) = \sums w_s \wt\bmu_\s=\wt\K\left(\w,\wt\bb\right)\y.\notag
	\ee
	The $J$-fold CV criterion can be obtained by
	$CV_{J_n}(\w) = \left\|\wt\bmu(\w)-\y\right\|^2$,
	and the $J$-fold CV choice of the weight vector is the value that minimizes $CV_{J_n}(\w)$ over $\w\in\calW$, i.e.,
	\be\label{eq:optw}
	\wh\w = \argmin_{\w\in\calW} CV_{J_n}(\w).
	\ee
	The resulting averaging estimator of $\bmu$ is $$\wh\bmu(\wh\w)=\sums \wh w_s\wh\bmu_\s=\K(\wh\w,\wh\bb)\y,$$
	which we refer to as the $J$-fold CV model-averaging (JCVMA) estimator.
	
	Since the CV objective function can be rewritten as $CV_{J_n}(\w)=\w\tt\A\w$, where the element of $\A$ is $A_{s,m}=\{\K_\s(\wt\bb_\s)\y-\y\}\tt\{\K_{(m)}(\wt\bb_{(m)})\y-\y\}$ for $s,m=1,2,\ldots,S_n$, the weight calculation in~\eqref{eq:optw} is a quadratic programming problem which is typically easy to solve. The computational cost in this optimization mainly lies in the CV estimator of $\{\wt\bmu_\s\}_{s=1}^{S_n}$, and can be substantial if $S_n$ and $J_n$ are large.
	{We shall discuss how to determine $S_n$ and which models to combine when the entire model space is huge in Section~\ref{sec:pbigern}.}
	
	\section{Asymptotic properties}\label{sec:asp_prop}
	This section studies the asymptotic properties of the proposed averaging estimator. With the goal of prediction, we first examine the squared loss of our averaging estimator when all candidate models are approximations of the DGP and are consequently misspecified. We show that {in this case,} the $J$-fold CV weighting is asymptotically optimal in the sense that the resulting squared loss is asymptotically identical to that from the infeasible best possible averaging estimator. Second, we consider that the set of candidate models includes the correct (but not necessarily true) models.
	A model is correct if it nests the DGP, i.e., the $s\th$ model is correct if and only if there exists a vector $\bbeta_\s$ such that
	$\mu_i=
	g_\s(\x_{\s,i}\tt\bbeta_\s)$.
	%
	%
	%
	Thus, the correct model is not unique.
	In contrast, if some covariates in the DGP are omitted by a candidate model, this candidate model is called a misspecified model. The correct models exist if the DGP is a single-index model with covariates being a subset of all given covariates.
	In this case, we establish the consistency of selecting the correct models, i.e., the weights assigned to the correct models approach 1 as the sample size increases.
	
	\subsection{Asymptotic optimality}
	\label{sec:Asymptotic}
	
	This section studies the property of JCVMA when none of the candidate models is correct. The deviation between the true and candidate models can be in the specification of the single-index structure and/or the covariates.

	The following regularity conditions are required for the asymptotic optimality of the JCVMA estimator. All limiting processes below correspond to $n\to \infty$ unless stated otherwise.
	
	\begin{asu}\label{con:qusitrue}
		For $s=1,2,\ldots,S_n$ and $r=1,2,\ldots,p_s$, the $r\th$ element of $\hbeta_\s$ obtained from \eqref{eq:objective},
		$\hat{\beta}_{\s,r}$, has a limiting value $\beta^\ast_{\s,r}$. 
	\end{asu}
	This condition ensures the existence of the limit of slope parameters which are often referred to as ``quasi-true'' parameters \citep[Theorem 3.2 of][]{White1982}, and similar conditions are imposed in many model averaging studies, such as \citet{Xinyu2016Optimal}, \citet{Ando2017}, among others.
	%
	%
	%
	Further, we denote
	$\bmu^*_\s = \K_\s(\sbeta_\s)\y$ and $\wt\bmu^*_\s = \wt\K_\s({\bf 1
	}_{J_n}\otimes \sbeta_\s)\y$, where $\otimes$ denotes the Kronecker product and ${\bf 1}_{J_n}$ is a $J_n\times 1$ vector of 1.
	
	\begin{asu}\label{con:mu}
		\subasu \label{eq:C41}  $\sigma_{\max}=O(1)$, where $\sigma_{\max}=\supi\sigma_{i}$.
		\subasu  \label{eq:C42} $\supi|\mu_i|=O(1)$.
	\end{asu}
	Condition~\ref{con:mu} restricts the magnitude of the variance and the mean of $y_i$.
	Similar conditions are imposed by \cite{Ando2017} (Assumptions (A1) and (A4)) and \cite{ZhuA} (Conditions~(C.1) and (C.7)).

	\begin{asu}\label{con:kernel}
		There exists a positive sequence $\{d_n\}$ such that
		\begin{align}
		\sups\supij\frac{\kernel(\x_{\s,i}\tt\sbeta_\s-\x_{\s,j}\tt\sbeta_\s)}{\sumjsn\kernel(\x_{\s,i}\tt\sbeta_\s-\x_{\s,\js}\tt\sbeta_\s)}=O_P(d_n)
		\text{ and } d_nn=O(1).\label{eq:C211}
		\end{align}
	\end{asu}
	This condition requires sufficient variation in the covariates to explain $\bmu$. For example, if $\{\x_i\}$ is independent and identically distributed (i.i.d.), \eqref{eq:C211} holds with probability one. {For each fixed  candidate model, the sum of each row of $\K_{(s)}(\bbeta^*_{(s)})$ is one, so $d_n$ is often proportional to $n^{-1}$.} 
	\begin{asu}\label{con:Ksij}
		The smoother matrix satisfies
		\be
		\sups\supj\sumin\nolimits K_{\s,ij}(\sbeta_\s)=O_P(1).\notag\label{eq:C21}
		\ee
	\end{asu}
	Condition~\ref{con:Ksij} concerns the  $L_{\infty}$ norms of $\K_\s$ and is widely used in nonparametric models, such as 
	Assumption 1.3.3(i) of \cite{Hardle2000Partially}.

	\begin{lemma}\label{lem:qusitrue}\label{lem:consistance}
		{Under Conditions \ref{con:qusitrue}--\ref{con:Ksij} and \ref{con:conlast}--\ref{con:beta2supplement} in Appendix~\ref{sec:conditions}, we have that}
		\begin{align}
		\sups\sqrt{\frac{n}{S_n p_s}}\left\|\hbeta_\s-\sbeta_\s \right\|=O_P(1),\label{eq:lem1-1}
		\end{align}
		and
		\begin{align}
		\max_{1\leq j\leq J_n}\sups\sqrt{\frac{n-M_n}{S_n p_s}}\left\|\hbeta_\s^{[-j]}-\sbeta_\s \right\|=O_P(1).\label{eq:lem1-2}
		\end{align}
	\end{lemma}
	This lemma states that the NLS estimator $\hbeta_\s$ obtained from minimizing~\eqref{eq:objective} and its CV version $\hbeta_\s^{[-j]}$ (leaving observations of the $j\th$ block out) both converge to the quasi-true value $\sbeta_\s$ of the $s{\th}$ model at the uniform speed of $\sqrt{n/S_np_s}$ and $\sqrt{(n-M_n)/S_np_s}$, respectively. Importantly, these convergence results hold in both finite- and divergent-dimensional cases, which is vital for establishing the asymptotic optimality.

	To state the next condition regarding the second-order derivatives of the link function, we denote
	$$
	\hat{g}_\s(\x_{\s,i}\tt\bbeta_\s)=\sumjn\nolimits y_j \kernel(\x_{\s,i}\tt\bbeta_\s-\x_{\s,j}\tt\bbeta_\s)/\sumjsn\nolimits\kernel(\x_{\s,i}\tt\bbeta_\s-\x_{\s,\js}\tt\bbeta_\s)\notag
	$$
	and
	$$
	\hat{g}_\s^{[-\setB(i)]}(\x_{\s,i}\tt\bbeta_\s)=\sumjn\nolimits y_j \kernel(\x_{\s,i}\tt\bbeta_\s-\x_{\s,j}\tt\bbeta_\s)/\sum\nolimits_{\js\in\setA(i)}\kernel(\x_{\s,i}\tt\bbeta_\s-\x_{\s,\js}\tt\bbeta_\s),\notag
	$$
	where the superscript $[-\setB(i)]$ denotes the estimator without using the entire block that contains the $i\th$ observation, i.e.,
	\begin{align}
	\setB(i)=\left\{\lceil i/M_n\rceil M_n-M_n+1,\lceil i/M_n\rceil M_n-M_n+2,\ldots,\lceil i/M_n\rceil M_n\right\},\label{eq:calB}
	\end{align}
	$\lceil \cdot \rceil$ is the ceiling of a number, and $\setA(i)=\{1,2,\ldots,n\}\backslash\setB(i)$.
	We further denote $\mathcal{O}(\sbeta_\s,\rho)$ as a neighborhood of $\sbeta_\s$ for some positive constant $\rho$, i.e., $\{\bbeta_\s\in \mathcal{R}^{p_s}: \|\bbeta_\s-\sbeta_\s\|\leq\rho\}$, and $\lambda_{\max}(\cdot)$ as the maximum eigenvalue.


	\begin{asu}\label{con:lambda}
		There exists a $\rho>0$ such that
		\begin{align}
		\sups \sup_{\bbeta_\s^{(1)},..,\bbeta_\s^{(n)}\in \mathcal{O}(\sbeta_\s,\rho)}\lmax\left\{\onen\sumin \frac{\partial \hat{g}_\s(\x_{\s,i}\tt\bbeta_\s^{(i)})}{\partial \bbeta_\s} \frac{\partial \hat{g}_\s(\x_{\s,i}\tt\bbeta_\s^{(i)})}{\partial \bbeta\tt_\s} \tt\right\}=O_P(p_{\max}),\label{eq:lambda}
		\end{align}
		and
		\begin{align}
		\sups \sup_{\bbeta_\s^{(1)},..,\bbeta_\s^{(n)}\in \mathcal{O}(\sbeta_\s,\rho)}\lmax\left\{\onen\sumin\frac{\partial \hat{g}_\s^{[-\setB(i)]}(\x_{\s,i}\tt\bbeta_\s^{(i)})}{\partial \bbeta_\s} \frac{\partial \hat{g}_\s^{[-\setB(i)]}(\x_{\s,i}\tt\bbeta_\s^{(i)})}{\partial \bbeta\tt_\s}\right\}=O_P(p_{\max}),\label{eq:lambda2}
		\end{align}
		where $p_{\max}=\max_{1\leq s\leq S_n} p_s$ denotes the maximum dimension of candidate models.
	\end{asu}

	In this condition, \eqref{eq:lambda} essentially controls the magnitude of $\|\hat{\bmu}_\s-\bmu^*_\s\|^2$ through the differential mean value theorem, and \eqref{eq:lambda2} is a CV version of~\eqref{eq:lambda} due to the $J$-fold CV estimator, which controls the magnitude of $\|\wt\bmu_\s-\wt\bmu^*_\s\|^2$.
	%
	This condition is satisfied if $\{g_\s(\cdot)\}_{s=1}^{S_n}$ is sufficiently smooth, such as $\sups\|\partial g_\s(\x_\s\tt\bbeta^*_\s)/\partial \bbeta^*_\s\|^2=O(p_{\max})$ for each {$s=1,2,\ldots,S_n$}.
	Similar conditions are often used {when studying model averaging for parametric models}, e.g., Condition (C.4) of \cite{Xinyu2016Optimal} and Condition~(C.2) of \cite{zhang2019Parsimonious}.

	%
	%
	
	Denote $\bmu^\ast(\w)=\sumsn w_s \K_\s(\bbeta_\s^\ast)\y$ as the averaging estimator based on the quasi-true parameters {$\bbeta_{(s)}^\ast$ for $s=1,\ldots,S_n$}. Denote $L_n^\ast(\w)= \left\|\bmu^\ast(\w)-\bmu\right\|^2$ as the corresponding squared loss, and $\xi_n=\inf\nolimits_{\w\in \calW}L_n^\ast(\w)$ as the minimum squared loss over all averaging estimators. We assume the following condition.
	\begin{asu}\label{con:xi}
		\subasu\label{eq:C61}
		$\xi_n^{-1}S_n^{1/2}np_{\max}(n-M_n)^{-1/2}=o_P(1)$.\subasu\label{eq:C62}
		$\xi_n^{-1}d_n M_n n =o_P(1)$.
	\end{asu}
	This set of conditions resembles (8) in Theorem 1' of \cite{wan2010least} and Condition~(C.2) of \cite{ZhuA}, which {essentially requires that all candidate models are misspecified to a non-trivial extent, such that their mean squared errors are not too small. Thus, it precludes any scenario in which the correct models are included in the set of candidate models.} To better understand this condition, {
		consider a situation where the set of candidate models includes correct models. In this case, if we denote $s_0$ as the index of a correct model, then $\mu_i=
		g_{(s_0)}(\x_{(s_0),i}\tt\sbeta_{(s_0)})$, and
\begin{align*}
\xi_n&
=\inf_{\w \in\calW}\left\|\bmu^\ast(\w)-\bmu\right\|^2\leq\left\|\K_{(s_0)}(\bbeta_{(s_0)}^\ast)\y-\bmu\right\|^2=\sumin\left\{\hat{g}_{(s_0)}(\x_{\s,i}\tt\sbeta_{(s_0)})-\mu_i\right\}^2\notag\\
&=n\left\{O_P(h^2_{s_0}+n^{-1/2}h^{-1/2}_{s_0})\right\}^2=O_P\left(nh^4_{s_0}+h_{s_0}^{-1}\right),
\end{align*}
where $h_{s_0}$ is the bandwidth for the $s_0\th$ model and the last equality can be deduced from the asymptotic distribution of semiparametric estimators under some regularity conditions \citep[see, e.g.,][Theorem 5.3 and Theorem 6.4]{Fan1996Local}. Because $S_n^{1/2}n^{1/2}p_{\max}/(nh_{s_0}^4+h_{s_0}^{-1})$ usually does not converge to 0 {with general restrictions on $h_{s_0}$}, Condition~\ref{eq:C61} is violated.}
	{Note that this condition does not conflict with Condition \ref{con:lambda} because {they concern different distance measures. In particular,} Condition~\ref{con:lambda} controls the distance {between the estimators of $\bmu_\s$ (i.e., $\wh\bmu_\s$ and $\wt\bmu_\s$) and their corresponding quasi-true values (i.e., $\bmu_\s^\ast$ and $\wt\bmu_\s^\ast$), while Condition~\ref{con:xi} concerns the degree of misspecification which is the distance between the quasi-true value $\bmu_\s^\ast$ and the true value $\bmu$.}}
	
	
	{Moreover, Condition~\ref{con:xi} also provides restrictions on the relative divergent rates of $p_{\max}, S_n, M_n$ and $\xi_n$: namely, $\xi_n$ is required to grow at a rate no slower than $S_n^{1/2}np_{\max}(n-M_n)^{-1/2}$ and $d_n M_n n$. For example, if $\xi_n$ explodes at a rate of $n^{1-\alpha}$ for some $\alpha>0$, then $S_n^{1/2}p_{\max}(n-M_n)^{-1/2}n^\alpha$ and $d_n M_n n^\alpha$ are both required to converge to 0, which further implies that $\alpha$ needs to be small. 
		If $\xi_n$ explodes at a rate of $n$, $S_n=O(1)$ and $M_n=O(1)$, then we can allow $p_{\max}$ to grow at a rate of $n^{1/2-c}$ for some positive constant $c<1/2$}. 
	Overall, Condition~\ref{con:xi} is more likely to be satisfied when $\xi_n$ approaches infinity at a faster rate, or in other words, when all candidate models are misspecified to a larger extent such that the squared loss of the best possible averaging estimator is large.

	\begin{Th} \label{thm:opt}
		{Under the conditions of Lemma~\ref{lem:qusitrue} and Conditions \ref{con:lambda}--\ref{con:xi}}, we have that
		\be
		{L_n(\wh\w) \over
			\inf_{\w\in \calW }L_n(\w)}\to 1\notag\quad \textrm{in probability}. \label{opt1}
		\ee
		
	\end{Th}
	Theorem~\ref{thm:opt} shows that the JCVMA estimator of $\bmu$ is asymptotically optimal in the sense that it leads to a squared loss that is asymptotically identical to that of the infeasible best possible model-averaging estimator.
	%
	
	\subsection{Consistency of averaging weights}
	This section studies the limiting behavior of averaging weights when the set of candidate models includes at least one correct model. In this case, we wish to find the correct model(s) which enables us to build the relationship between the covariates and the response variable and provides better prediction than using misspecified models under certain conditions, as specified in the details below.

	Without loss of generality, we assume that the first $S_0$ ($\geq 1$) models are correct.
	We denote $\wh w_\Delta=\sumsz \wh w_s$ as the sum of weights given to the $S_0$ correct models, where $\wh w_s$ is the $s\th$ element of the JCVMA weight vector $\wh\w$.
	Denote $\mathcal{W}_F=\left\{\w\in \calW:w_s=0\text{ for }s=1,2,\ldots,S_0\right\}$ as the set of weight vectors that assign zero weights to the correct models. Let $\xi_F=\inf_{\w\in \mathcal{W}_F}L_n^\ast(\w)$ be the squared loss of only averaging misspecified models. To obtain the limiting property of the averaging weights, an extra condition is needed.
	\begin{asu}\label{con:C7}\subasu\label{eq:C71}
		$\xi_F^{-1}S_n^{1/2}np_{\max}(n-M_n)^{-1/2}=o_P(1)$.
		\subasu\label{eq:C72} $\xi_F^{-1}d_n M_n n =o_P(1)$.
	\end{asu}
	Condition~\ref{con:C7} requires that the squared loss of the best possible averaging of misspecified models has a sufficiently large divergent rate, in order to distinguish between the misspecified and correct models. Such requirements are similar to Condition~\ref{con:xi}
with $\xi_n$
replaced by $\xi_F$.
%
%
Condition~\ref{con:C7} is a counterpart of Condition~\ref{con:xi} for the cases in which correct models are present in the model space. It restricts how the misspecified models deviate from the true one as well as the relative divergent rates of $S_n$, $d_n$, $p_{\max}$ and $M_n$ when correct models exist. A similar set of conditions is discussed for linear regressions in~\cite{zhang2019Parsimonious}.

		The next theorem demonstrates the asymptotic behavior of averaging weights when correct models are included in the set of candidate models.
		\begin{Th}\label{thm:consist}
			{Under the conditions of Lemma~\ref{lem:consistance} and Conditions \ref{con:lambda} and~\ref{con:C7}}, we have that $\wh w_\Delta\to 1$ in probability.
		\end{Th}
		Theorem~\ref{thm:consist} shows that JCVMA tends to assign all weights to the correct models if they exist in the candidate model set. Consistent selection of the correct models enables us to examine the (nonlinear) relation between covariates and the response variable.
		
		To conclude the prediction performance when candidate models include the correct models, we need the following extra condition for $\xi_F$, $S_n$ and $p_{\max}$.

		\begin{asu}\label{con:C10}
\subasu \label{eq:C101} $\xi_F^{-1} S_np^2_{\max}=o_P(1)$. \subasu\label{eq:C102}
			$\xi_F^{-3}\left\{S_n^{1/2}n^3p_{\max}(n-M_n)^{-1/2}+d_n M_n n^3\right\}=o_P(1)$.
		\end{asu}
Condition~\ref{eq:C101} imposes a stronger restriction on the speed that $S_n$ and $p_{\max}$ diverge.
		{Condition~\ref{eq:C102} adds the two equalities in Condition~\ref{con:C7}, and multiplies the left-hand side by $\xi_F^{-2}n^2$.} Note that the limit of $\xi_F^{-2}n^2$ is usually not zero, so $\xi_F^{-1}S_n^{1/2}np_{\max}(n-M_n)^{-1/2}$ and $\xi_F^{-1}d_n M_n n$ must converge to zero faster than those in~Condition \ref{con:C7}. More specifically, combining  Condition~\ref{eq:C102} with the fact that $\xi_F=O_P(n)$, a result implied by the expression of $L_n^*(\w)$, we have that
		\begin{align*}
		&\quad\xi_F^{-1}\left\{S_n^{1/2}np_{\max}(n-M_n)^{-1/2}+d_n M_n n \right\}\\
		&=\xi_F^2 n^{-2} \xi_F^{-3}\left\{S_n^{1/2}n^3p_{\max}(n-M_n)^{-1/2}+d_n M_n n^3 \right\}=o_P(1),
		\end{align*}
		which further implies Condition~\ref{con:C7}.
		%

\begin{corollary}\label{thm:opt-correctmodel}
Under the conditions of Lemma~\ref{lem:consistance} and Conditions~\ref{con:lambda} and \ref{con:C10}, we have that
\begin{align}
\frac{L_n(\hat{\w})}{\inf_{\w\in\mathcal{W}_F}L_n(\w)}\to 0\quad \textrm{in probability}.\notag
\end{align}
\end{corollary}
This corollary establishes the asymptotic optimality when correct models are contained in the candidate model set, which complements the asymptotic optimality in Theorem~\ref{thm:opt}. It shows that when correct models are available in the candidate set, the squared loss of the JCVMA estimator of $\bmu$, namely, $L_n(\hat{\w})$, {is asymptotically negligible compared to that of any averaging estimator that assigns zero weights to the correct models.} In conjunction with Theorem~\ref{thm:consist}, this corollary suggests that JCVMA also provides good prediction when correct models are available, since it asymptotically assigns all weights to the correct models and outperforms any other averaging predictions that fail to include these models. {Note that the asymptotic optimality in Corollary~\ref{thm:opt-correctmodel} concerns the squared loss of the averaging estimator of $\bmu$ but does not directly suggest the estimation efficiency of $\bbeta$. Thus, it differs from the semiparametric efficiency bounds studied in \citet{Ichimura1993Semiparametric}.
		}

\section{Model averaging based on regularized estimation and pre-screening}\label{sec:pbigern}
{Thus far, we have studied SIM averaging when $p_{\max}<n$ and $S_n$ is not too large, even though both of them are allowed to diverge as $n$ increases. In some applications, there may exist a huge number of potential covariates such that some candidate models have more parameters to estimate than the sample size and the  number of all possible models is overly large. Hence, in this section, we study how to perform JCVMA in such situations. We first consider averaging regularized estimators for candidate models in the presence of many covariates, and then study how to choose $S_n$ and the set of candidate models when the entire model space is too large to be completely considered.
	
	\subsection{Averaging regularized estimators}
	When there exist a large number of covariates, NLS estimators obtained from solving~\eqref{eq:objective} can be rather inefficient and sometimes even infeasible for some candidate models due to (too) many parameters. Hence, we consider an alternative method to estimate the $s\th$ candidate SIM using NLS with an $L_1$ penalty. Particularly, the estimator of $\bbeta_\s$ for the $s{\th}$ candidate model can be obtained as
	\begin{align}
	\hbeta_\s^R=\arg\min_{\bbeta_\s}\left\{H_{\s,n}(\bbeta_\s)+\lambda_s \|\bbeta_\s\|_1\right\},\label{eq:high_obj}
	\end{align}
	where $H_{\s,n}(\bbeta_\s)$ is the NLS objective function of the $s\th$ model defined in~\eqref{eq:objective}, $\|\bbeta_\s\|_1=\sum_{i=1}^{p_s} |\beta_{i}|$ is the penalty, and  $\lambda_s$ is the model-specific tuning parameter.
	The above optimization problem can be solved, e.g., by the coordinate descent algorithm \citep{Friedman2010Regularization}.
	Furthermore, we denote $\wh\bmu^R_\s=\K_\s(\hhbeta_\s)\y, \wh\bmu^R(\w)=\sums w_s\wh\bmu^R_\s$ and $L^R_n(\w)=\|\wh\bmu^R(\w)-\bmu\|^2$.
	
	To study the property of the regularization-based JCVMA estimator, we need to impose the conditions to ensure the consistency of regularized estimators, and also restrict the relative divergent speed of $S_n$ and $d_n$, parallel to Lemma~\ref{lem:qusitrue} and Condition~\ref{con:xi}, respectively, for the unregularized cases; see Section~S.1.5 in the Online Supplement  
	for more detailed discussions regarding these conditions. 
	\begin{corollary}\label{thm:high}
		Under Conditions \ref{con:mu}--\ref{con:Ksij} and
		\ref{con:high}--\ref{con:Hlambda} in Appendix~\ref{sec:conditions-col2}, we have that
		\be
		{L^R_n(\wh\w) \over
			\inf_{\w\in \calW }L^R_n(\w)}\to 1\notag\quad \textrm{in probability}. \label{opt1}
		\ee
	\end{corollary}
	Corollary~\ref{thm:high} shows that the asymptotic optimality of JCVMA continues to hold when the candidate SIMs are estimated by NLS with an $L_1$ penalty. This is the first optimality result of averaging regularized estimators.
	In conjunction with model screening discussed in the following subsection, the regularization technique offers a way to implement model averaging when the number of covariates exceeds the sample size.

	
	\subsection{Model averaging based on pre-screening}\label{sec:model-screening}
	When $p$ is particularly large or even exceeds the sample size, not only are some candidate models difficult to estimate, but the model space is also huge, rendering estimation and combination of all possible models infeasible. In this case, we can implement a model-screening step prior to averaging, which we refer to as pre-screening. Pre-screening can be used when $p<n$ but all possible combinations of covariates still lead to excessively numerous candidate models, i.e., $2^p$ is large, and it is also {useful in the high-dimensional cases} in which $p>n$.
	We propose two approaches to 
	{
		pre-screen the models} and construct the set of candidate models for averaging, depending on the relation between $p$ and $n$.
	
	First, when $p<n$ and estimating the full model with all covariates is feasible, we can order the covariates based on their marginal correlations with the response variable, and construct the set of candidate models by including one extra covariate at each time based on the ordering. The idea of model screening based on bivariate correlation is in a similar spirit as that of the ``sure independence screening'' proposed by \citet{fan&lv:2008}. Similar screening procedures have been used in other model-averaging studies, such as \cite{claeskens.croux.ea:2006} and \cite{Ando2014}.

	Second, when $p>n$, it is impossible to estimate the full model using the standard NLS as in~\eqref{eq:objective}, and we propose to pre-screen the models based on regularized estimation of the full model with certain tuning parameters as in~\eqref{eq:high_obj}. {Particularly, we can solve the following optimization problem: $\min_{\bbeta}\left\{H_{n}(\bbeta)+{\lambda} \|\bbeta\|_1\right\}$, where $H_{n}(\bbeta)$ is the same objective function as~\eqref{eq:objective} but using all the covariates, {and $\lambda$ is the tuning parameter}. With a feasible amount of different values of the tuning parameter $\lambda$, we can obtain {a set of} corresponding candidate estimators, which can then be conveniently averaged. We shall discuss how to choose a set of tuning parameters in practice in Section~\ref{sec:sub54}.}
	The idea of using regularized estimation for screening is advocated by \cite{Xinyu2016Optimal}, but they only consider the \emph{parametric} candidate models.

	To justify the SIM averaging estimator obtained after a preliminary model screening step, we study whether it {remains asymptotically optimal, i.e., whether the squared loss of the post-screening JCVMA is asymptotically identical to that of the infeasible best possible model-averaging estimator obtained from the original model set $\calW$ (without pre-screening).
	}
	Let $\mathcal{D}$ be a (random) subset of $\{1,2,\ldots,S_n\}$ and $\mathcal{W}^{\mathcal{D}}=\{\w\in[0,1]^{S_n}: \sum_{s\in\mathcal{D}} w_s= 1\ \textrm{and}\ \sum_{s\notin\mathcal{D}} w_s= 0\}$ be a subset of $\calW$. Note that $\mathcal{W}^{\mathcal{D}}$ is also random due to the randomness of $\mathcal{D}$. The post-screening model-averaging estimator based on the subset $\mathcal{D}$ is obtained by using the weight vector $\hat{\w}^s=\arg\min_{\w\in \mathcal{W}^{\mathcal{D}}}\ CV_{J_n}(\w)$. We make an additional assumption that there exists a non-negative series of $\nu_{n}$ and a weight series of $\w_{n}\in\calW$, such that
	$\xi_{n}^{-1}\nu_{n}=o_P(1)$, $\inf_{\w\in\calW} CV_{J_n}(\w)=CV_{J_n}(\w_{n})-\nu_{n}$, and $P(\w_{n}\in\mathcal{W}^{\mathcal{D}})\to 1$ as $n\to\infty$.
	This assumption enures that there exists a weight in $\mathcal{W}^{\mathcal{D}}$ to achieve the minimal CV loss asymptotically. This is the same as Assumption 1 in \citet{Xinyu2016Optimal}, in which more explanations are provided.
Under this additional condition as well as the conditions of Theorem~\ref{thm:opt}, we can then use the same arguments as Theorem 3 of \citet{Xinyu2016Optimal} to show that the post-screening model-averaging estimator based on the candidate model set $\mathcal{W}_n^{\mathcal{D}}$ still achieves the asymptotic optimality, namely
${L_n(\hat{\w}^s)/\inf_{ \w\in \mathcal{W}}L_n(\w)}\to 1$
in probability.}
%
%

	\section{Simulation study}\label{sec:simulation}
	This section examines the finite-sample performance of JCVMA and compares it with the popular IC-based model selection and averaging methods. We also report the performance of the full model that includes all covariates as a baseline. We first consider benchmark designs of $p<n$ and then study the case of $p>n$.

	\subsection{Benchmark experimental designs}\label{sec:dgp}
	To verify the theory in Section~\ref{sec:asp_prop}, we consider two exemplifying nonlinear functions that associate the response variable and covariates. For each nonlinear function, we study two cases that differ in the dimension of covariates. First, we fix the dimension of covariates to be finite. Second, we allow the dimension of covariates and the number of candidate models to be divergent. Furthermore, for each of the cases, we consider whether the correct models are included in the set of candidate models.
	
	\noindent\emph{Example 1:}
	We follow \citet{naik:tsai:2001} to consider the following DGP
	\begin{align*}
	y_i&=\mu_i+c\epsilon_i=\sin(\pi\x_i\tt\bbeta/6)+c\epsilon_i,\quad i=1,2,\ldots,n,
	\end{align*}
	where $\x_i$ is a $p\times 1$ vector generated from a multivariate normal distribution with mean zero and covariance matrix $\bSigma=(0.5^{|i-j|})_{p\times p}$. The settings of $p$ and $\bbeta$ vary across the four situations as specified below. $\epsilon_i$ is i.i.d.~and follows a standard normal distribution. $c$ controls the signal-to-noise ratio, and we vary $c$ such that $R^2=\var(\mu_i)/\var(y_i)$ ranges from 0.1 to 0.9.
	
	\noindent \emph{Example 2:} We follow \citet{kong.xia:2007} and~\citet{Ichimura1993Semiparametric} to consider the Tobit DGP as
	\begin{align*}
	y_i=(\mu_i+c\epsilon_i)I(\mu_i+c\epsilon_i>0)=(\x_i\tt\bbeta+c\epsilon_i)I(\x_i\tt\bbeta+c\epsilon_i>0),\quad i=1,2,\ldots,n,
	\end{align*}
	where $I(\cdot)$ is an indicator function, and the remaining settings are the same as Example 1.

	For each nonlinear link function, we consider the following four situations.
	\begin{itemize}
		\item[(1)] \textbf{Finite dimension with all candidate models misspecified}	
		
		We fix $p=7$ and set the coefficient vector as $\bbeta=(1, 1.5, 1, 0, 0.1, -1.5, 1.5)\tt$. To construct misspecified candidate models, we include the first covariate but omit the last in all candidate models. The remaining covariates are uncertain, but at least one of them is included, which leads to $S_n=2^5-1=31$ candidate models.
		
		\item[(2)] \textbf{Finite dimension with correct candidate models}	
		
		In this case, we set {$\bbeta=(1, 1.5, 0,1, 0, -1.5, 1.5)\tt$}, where two of the coefficients are set to zero to increase the number of correct models for demonstration purposes. All candidate models include the first and last covariates but differ in the specification of the remainders, so correct models are contained in the candidate model set.
		
		\item[(3)] \textbf{Divergent dimension with all candidate models misspecified}	
		
		To mimic the cases where the dimension of covariates increases with the sample size $n$, we set $\bbeta=(1,( \underline{1.5, 1, 0, 0.1, -1.5},\underline{1.5,1,0,0.1,-1.5},\ldots)_{\lceil 1.5n^{1/3}\rceil},1,1.5)\tt$, where the subscript $\lceil 1.5n^{1/3}\rceil$ is the speed at which the dimension of $\bbeta$ increases and $\lceil \cdot\rceil$ ascertains the ceiling of a number. We include the first covariate but omit the last two in all candidate models to construct misspecified candidate models as above. {To reduce the computational burden, we employ the pre-screening method based on an ordering of covariates as discussed in Section~\ref{sec:pbigern}, such that the number of candidate models also increases at a rate of $\lceil 1.5 n^{1/3}\rceil$. }
		
		\item[(4)] \textbf{Divergent dimension with correct candidate models}	
		
		The setting is similar to (3), except that we include the first and last two covariates in all candidate models and set $$\bbeta=(1,( \underline{1.5, 0, 1, 0, 0, -1.5, 0}, \underline{1.5, 0, 1, 0, 0, -1.5, 0},\ldots)_{\lceil1.5 n^{1/3}\rceil},1,1.5)\tt.$$		
	\end{itemize}We consider the sample sizes for estimation as $n=100$, 200, 300, 400 and 500, and set the testing size as 1,000; all results are reported based on $D=1000$ replications.

	\subsection{Implementation and comparison}\label{sec:Implement52}
	To implement the proposed {
		JCVMA}, we set the number of observations in each CV block to be $M_n=50$. Robustness checks suggest that the results are qualitatively similar as long as there are sufficient observations in each block. 
	{Following \citet{YuEmpirical}, we suggest to take the bandwidth of order $\kappa n^{-1/5}\log ^{-1/6}(n)$ and choose the optimal $\kappa$ via cross-validation.
		%
		We follow the convention to use the Gaussian kernel $K(u)=\exp(-u^2/2)/\sqrt{2\pi}$ when estimating each candidate model.}
	
	We compare {
		JCVMA} with three information criteria: AIC and BIC, and a variant of AIC, which is designed especially for SIMs.
	The AIC and BIC scores of the $s{\th}$ candidate model are given by
	\begin{equation*}
	\AIC_s=n\log(\hat{\sigma}^2_s)+2\tr\{\K_\s(\hbeta_\s)\},\quad
	\BIC_s=n\log(\hat{\sigma}^2_s)+\log(n)\tr\{\K_\s(\hbeta_\s)\},
	\end{equation*}
	where $\hat{\sigma}^2_s=n^{-1}\|\y-\hat{\bmu}_{\s}\|^2$.
	\cite{naik:tsai:2001} proposed a variant of AIC based on the Kullback-Leibler distance as
	\begin{align*}
	\AICC_s=\log(\hat{\sigma}_s^2)+\frac{n+\tr\left\{\hat{\H}_\s+\K_\s(\hbeta_\s)-\hat{\H}_\s\K_\s(\hbeta_\s)\right\}}{n-2-\tr\left\{\hat{\H}_\s+\K_\s(\hbeta_\s)-\hat{\H}_\s\K_\s(\hbeta_\s)\right\}},
	\end{align*}
	where $\hat{\H}_\s=\hat{\V}_\s(\hat{\V}_\s\tt\hat{\V}_\s)^{-1}\hat{\V}_\s\tt$ with
	\begin{align*}
	\hat{\V}_\s
	&=\left\{\partial \hat{g}_\s(\x_{\s,1}\tt\bbeta_\s)/\partial \bbeta_\s,\ldots,\partial \hat{g}_\s(\x_{\s,n}\tt\bbeta_\s)/\partial \bbeta_\s\right\}\tt \Big|_{\bbeta_\s=\hbeta_\s}\nonumber\\
	&=\left\{\hat{g}_\s^\prime(\x_{\s,1}\tt\hbeta_\s)\x_\s,\ldots,\hat{g}_\s^\prime(\x_{\s,n}\tt\hbeta_\s)\x_\s\right\}\tt,\end{align*}
	and $\hat{g}_\s^\prime(\cdot)$ denotes the derivative of $\hat{g}_\s(\cdot)$.
	
	We also compare the smoothed versions of the three information criteria, which use the values of the criteria for each candidate model as weights to construct the averaging estimators, namely, SAIC, SBIC and SAICC, e.g.,
	$
	\SAICC_s=\exp(-\AICC_s/2)/ \sum_{l=1}^{S_n}\exp(-\AICC_l/2).
	$
	
	We evaluate the performance of the methods from three perspectives. First, since our theory shows the asymptotic optimality of the {
		JCVMA}, we report the relative squared loss of each method with respect to the best possible averaging estimator, namely
	\begin{align*}
	D^{-1}\sum\nolimits_{d=1}^{D}L_n^{(d)}/\inf_{\w\in\calW} L_n^{(d)}(\w),
	\end{align*}
	where $\bmu^{(d)}$ is the true value of the testing set, $\hat{\bmu}^{(d)}$ is the predicted value produced by each method, $L_n^{(d)}=\|\hat{\bmu}^{(d)}-\bmu^{(d)}\|^2$ is the loss, $\inf_{\w\in\calW} L_n^{(d)}(\w)$ is the minimum squared loss over all possible averaging estimators, and the superscript $(d)$ denotes the $d{\th}$ replication.
	
	Second, we compare the prediction performance of various methods using the normalized mean squared prediction error (NMSPE), which is defined by
	$
	\NMSPE=D^{-1}\sum_{d=1}^{D}L_n^{(d)}/L^{(d)}_{\min},
	$
	where
	$L^{(d)}_{\min}$ is the minimum squared loss over all candidate models.
	
	Finally, to verify the consistency of weights when correct models exist in the candidate model set as shown in Theorem~\ref{thm:consist}, we plot the weights assigned to the correct models when $n$ increases. We also examine the validity of Corollary~\ref{thm:opt-correctmodel} by reporting the relative squared loss of JCVMA with respect to the best possible averaging estimators using only misspecified models, namely
	$
	D^{-1}\sum\nolimits_{d=1}^{D}L_n^{(d)}(\hat{\w})/\inf_{\w\in\calW_F} L_n^{(d)}(\w),
	$
	where $\hat\w$ is the JCVMA weight vector obtained by minimizing $CV_{J_n}(\w)$ as in~\eqref{eq:optw}.
	
	\subsection{Simulation results}
\begin{figure}[ht]
		\begin{center}
			\caption{Relative squared loss when all candidate models are misspecified ($R^2=0.5$)}		\label{fig:rsl-misspecified}
			\begin{tabular}{cc}
				\includegraphics[scale=0.7]{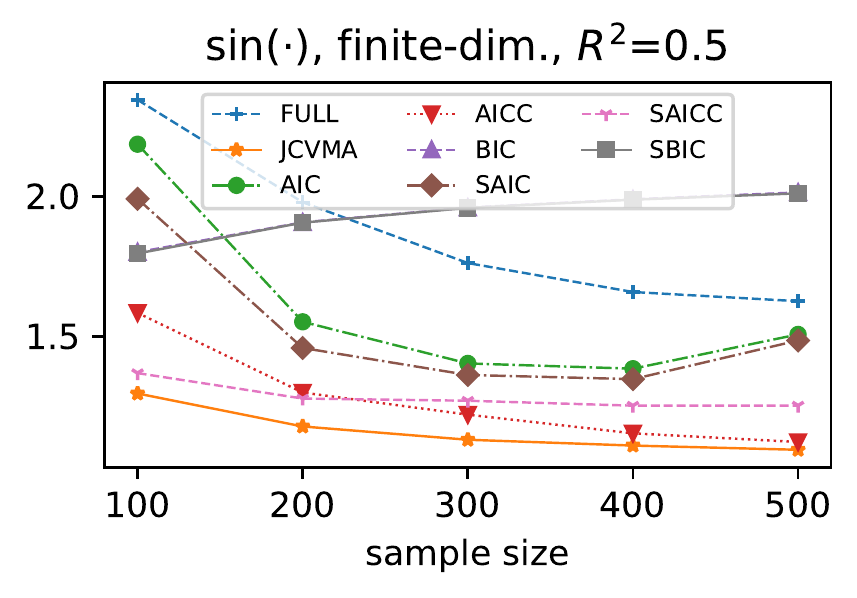}&
				\includegraphics[scale=0.7]{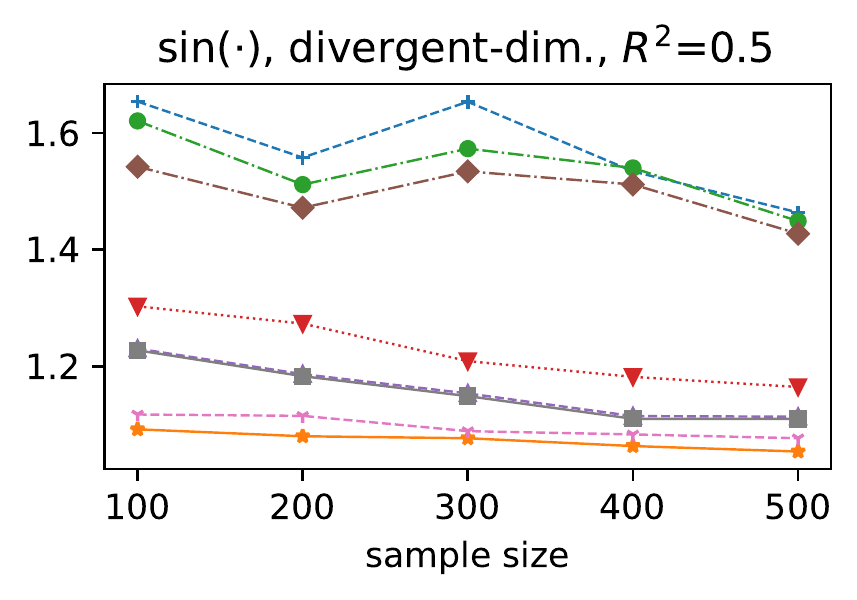}\\
				\includegraphics[scale=0.7]{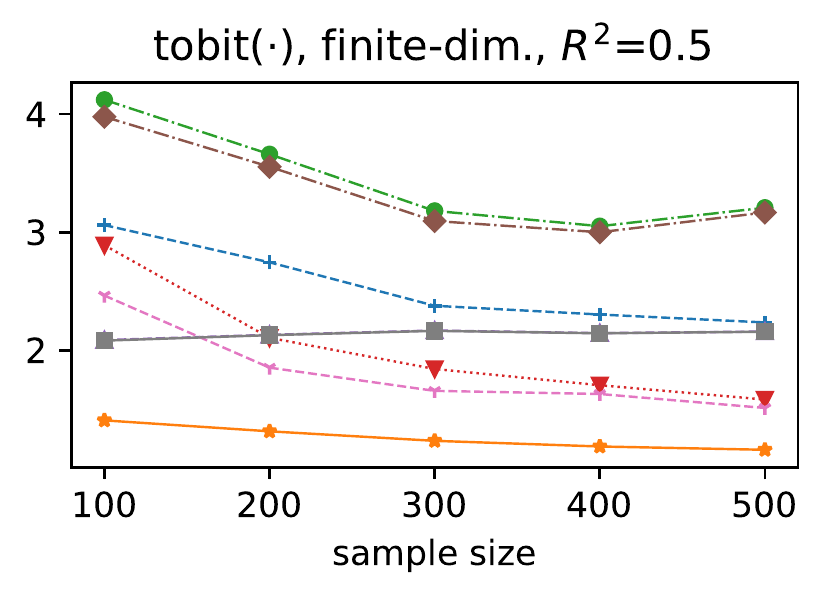}&
				\includegraphics[scale=0.7]{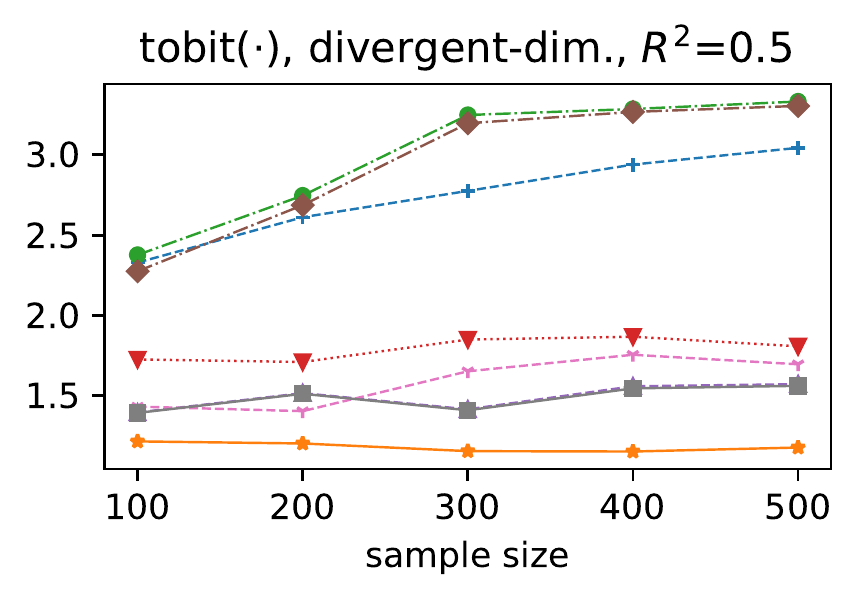}
			\end{tabular}
		\end{center}
		\footnotesize{\emph{Note:} This figure plots the relative squared loss defined as $L_n(\hat{\w})/\inf_{\w \in\calW}L_n(\w)$. The curves of BIC and SBIC largely coincide.}
	\end{figure}

	First, we examine the cases in which all candidate models are misspecified. To verify the asymptotic optimality in Theorem~\ref{thm:opt} and compare the performance of various methods, we present the relative squared loss with respect to the infeasible best possible averaging estimator in Figure~\ref{fig:rsl-misspecified}. To save space, we only report the results of $R^2=0.5$, which is closest to our empirical datasets. Increasing $R^2$ improves the performance of all methods, but the conclusion regarding the relative performance of all methods remains the same. The results of other levels of $R^2$ are provided in the Online Supplement. Figure~\ref{fig:rsl-misspecified} shows that the proposed JCVMA produces the lowest relative squared loss for both cases of fixed and divergent dimensions and for all sample sizes. Moreover, the relative squared loss of JCVMA generally decreases and tends to one when the sample size increases. The convergence of JCVMA confirms its asymptotic optimality as stated in Theorem~\ref{thm:opt}. In contrast, the curves of other averaging estimators do not show clear convergence to one.
	
\begin{figure}[htp]
		\begin{center}
			\caption{Normalized mean squared prediction error ($n=300$)}
			\label{fig:nmspe}		
			\small\begin{tabular}{cc}
				\multicolumn{2}{c}{All candidate models are misspecified}\\
				\includegraphics[scale=0.7]{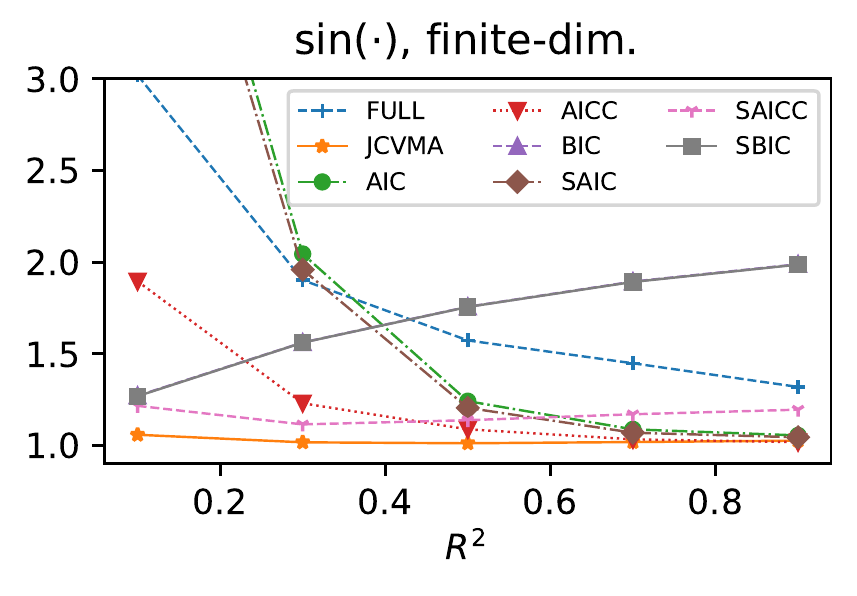}&
				\includegraphics[scale=0.7]{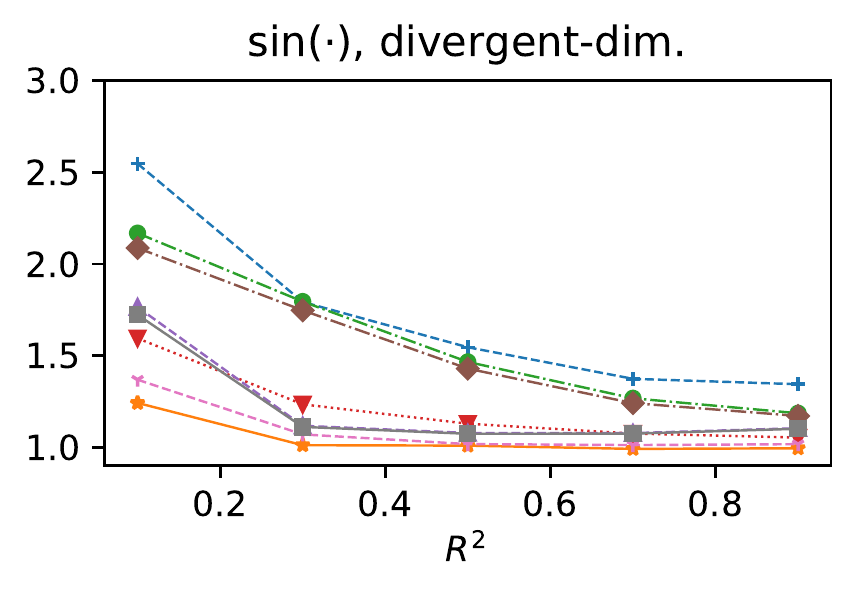}\\
				\includegraphics[scale=0.7]{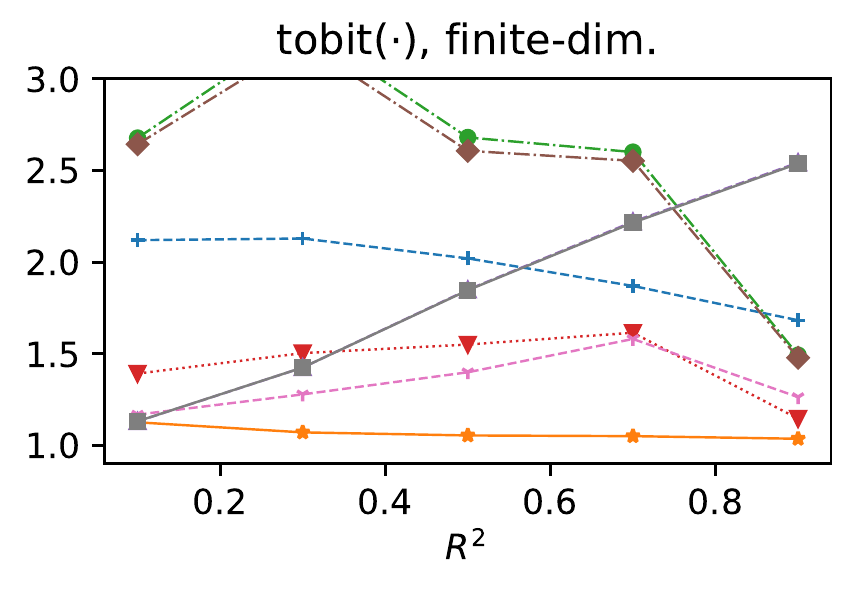}& 		
				\includegraphics[scale=0.7]{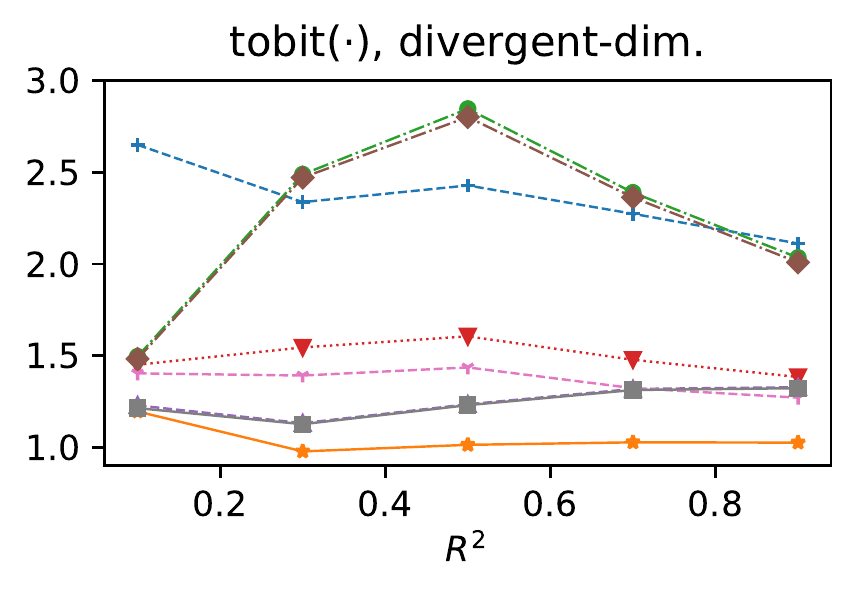}\\
				\multicolumn{2}{c}{Correct models are included in the candidate set}\\
				\includegraphics[scale=0.7]{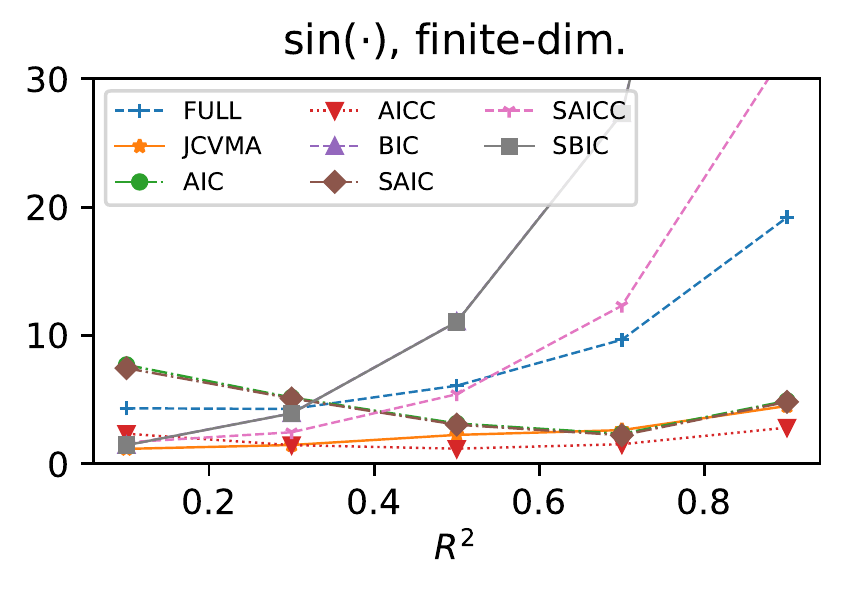}&
				\includegraphics[scale=0.7]{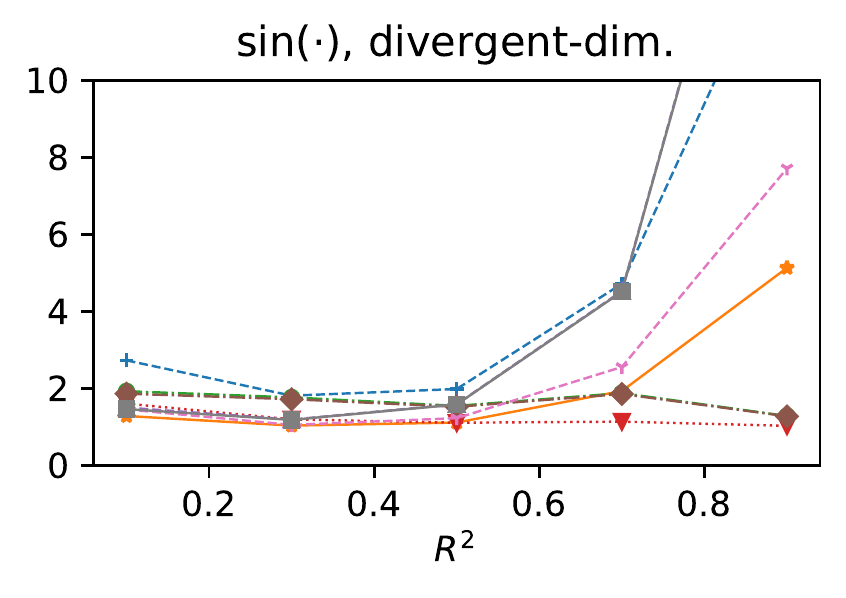}\\
				\includegraphics[scale=0.7]{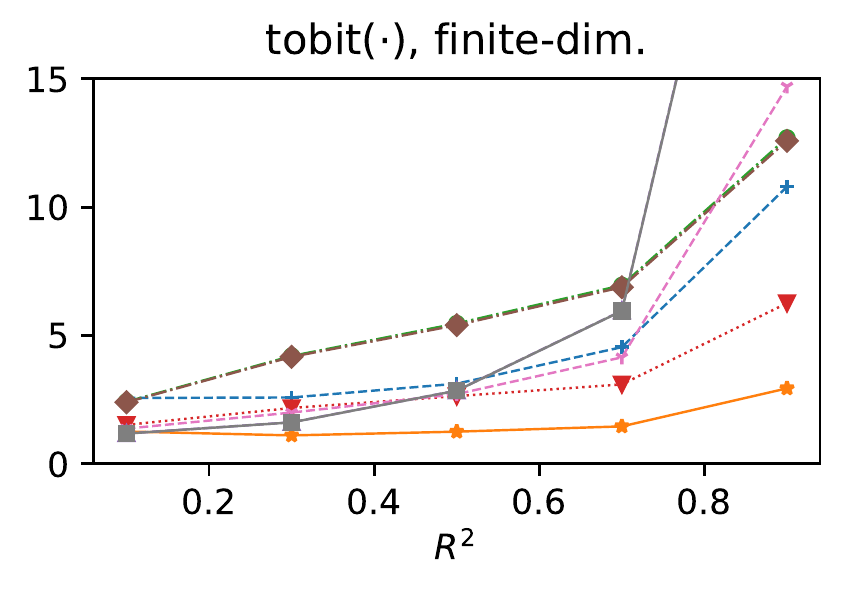}&
				\includegraphics[scale=0.7]{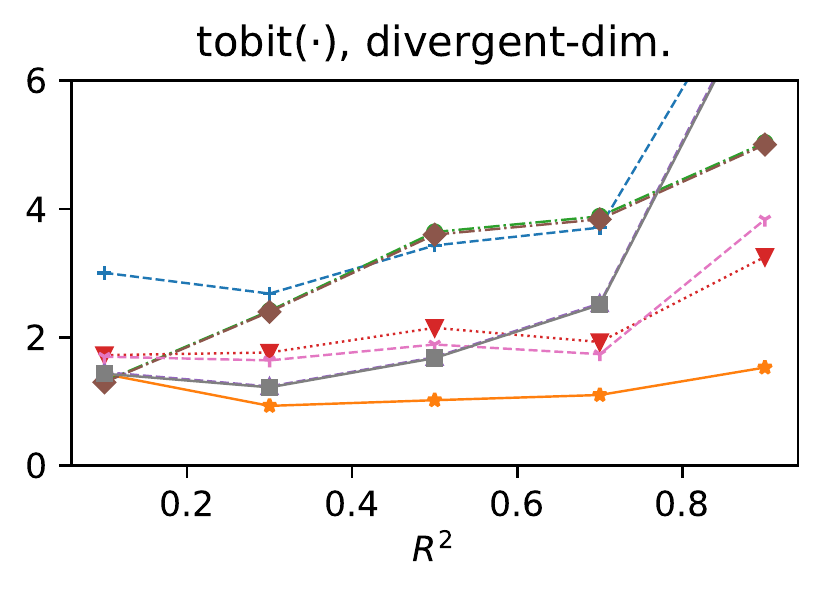}
			\end{tabular}
		\end{center}
		\footnotesize{\emph{Note:} The upper four figures plot the NMSPE when all candidate models are misspecified, and the bottom four figures consider the cases in which the candidate set includes correct models. The NMSPE is defined as $D^{-1}\sum_{d=1}^{D}L_n^{(d)}/L^{(d)}_{\min}$.}
	\end{figure}

	The upper four diagrams in Figure~\ref{fig:nmspe} compare the NMSPEs of the eight methods when all candidate models are misspecified. We report the results of $n=300$, while those of other sample sizes are highly similar and are thus provided in the Online Supplement. Again, we find that JCVMA produces the lowest NMSPE in almost all cases, followed by AICC or SAICC. The discrepancy between JCVMA and other methods seems larger in finite-dimensional cases than in divergent cases and appears to increase with $n$.

	Next, we consider the cases in which the candidate set includes correct models. The bottom four diagrams of Figure~\ref{fig:nmspe} present the related NMSPEs when $n=300$. In this case, the difference between the methods appears to be relatively small, especially when $R^2$ is small. Nevertheless, JCVMA continues to perform well with low NMSPEs and is always ranked among the top three methods, if not the best in a small proportion of cases. When $R^2$ is moderate or large, its improvement over other methods is particularly large for the Tobit model.
	
\begin{figure}[htp]
		\begin{center}		\caption{Properties of JCVMA when the candidate set includes correct models}
			\label{fig:weight&loss-correct}
			\begin{tabular}{cc}
				\multicolumn{2}{c}{Sum of weights assigned to correct models}\\
				\includegraphics[width=0.36\textwidth,height=0.19\textheight]{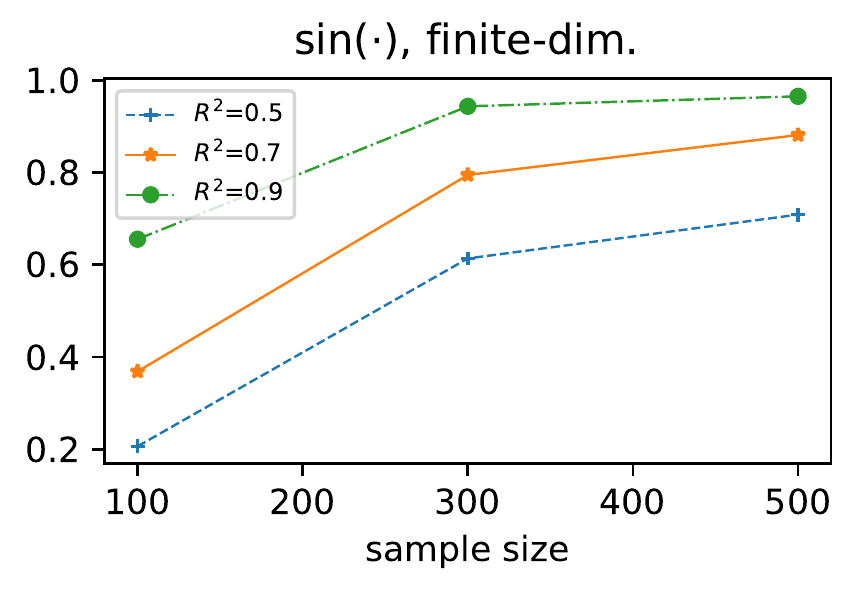}&
				\includegraphics[width=0.36\textwidth,height=0.19\textheight]{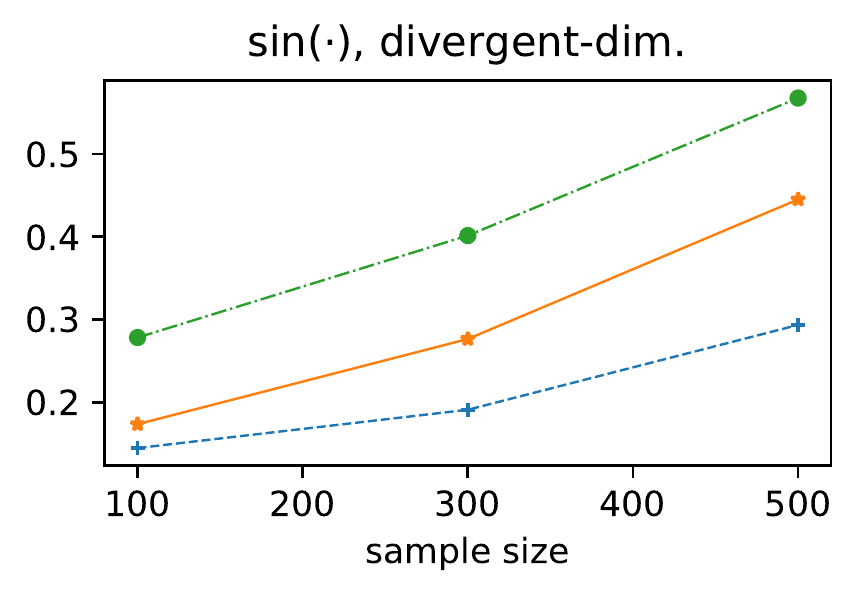}\\
				\includegraphics[width=0.36\textwidth,height=0.19\textheight]{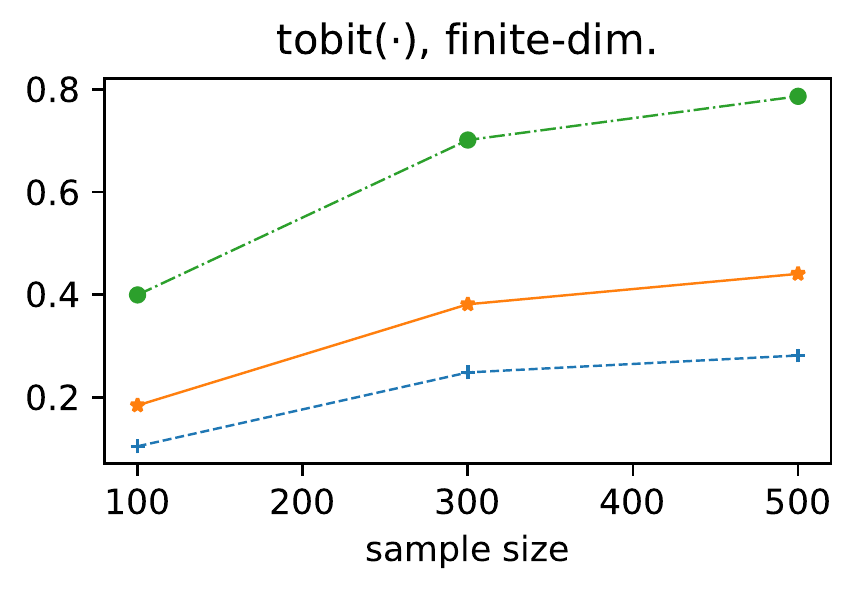}&
				\includegraphics[width=0.36\textwidth,height=0.19\textheight]{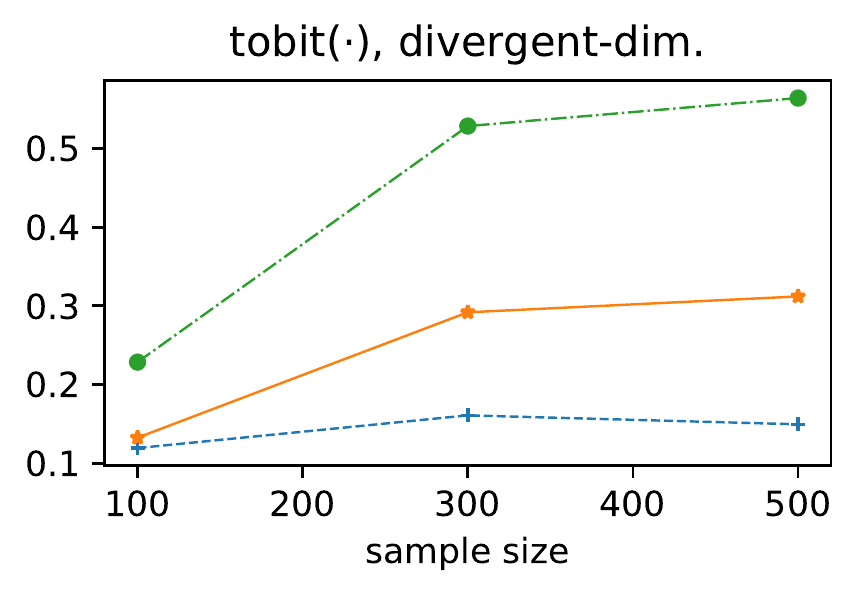}\\
				
				\multicolumn{2}{c}{Convergence of $L_n(\hat{\w})/\inf_{\w\in\calW_F}L_n(\w)$}\\
				\includegraphics[width=0.36\textwidth,height=0.19\textheight]{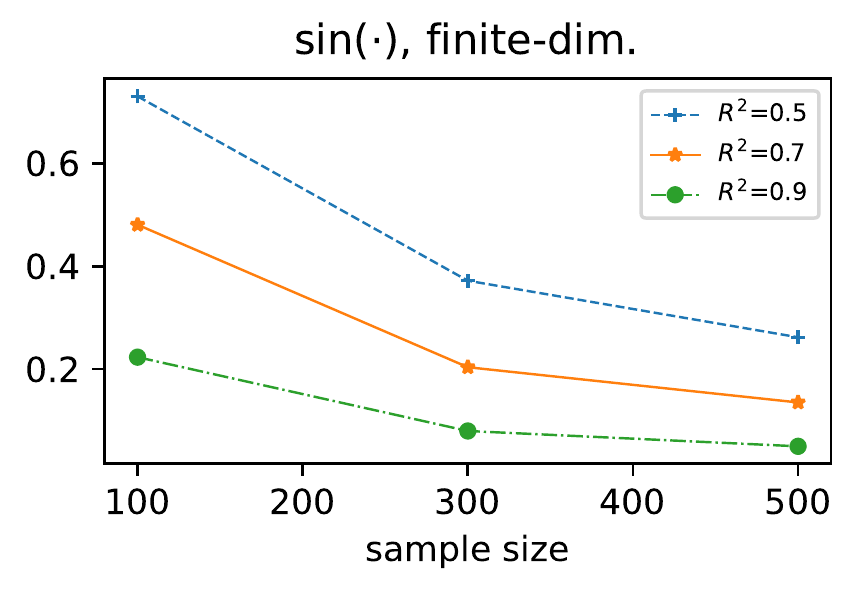}&
				\includegraphics[width=0.36\textwidth,height=0.19\textheight]{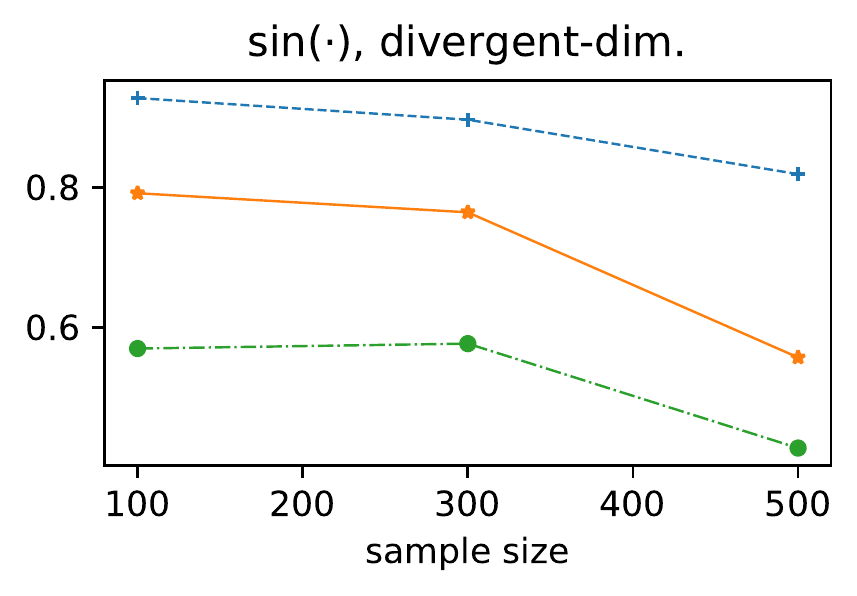}\\
				\includegraphics[width=0.36\textwidth,height=0.19\textheight]{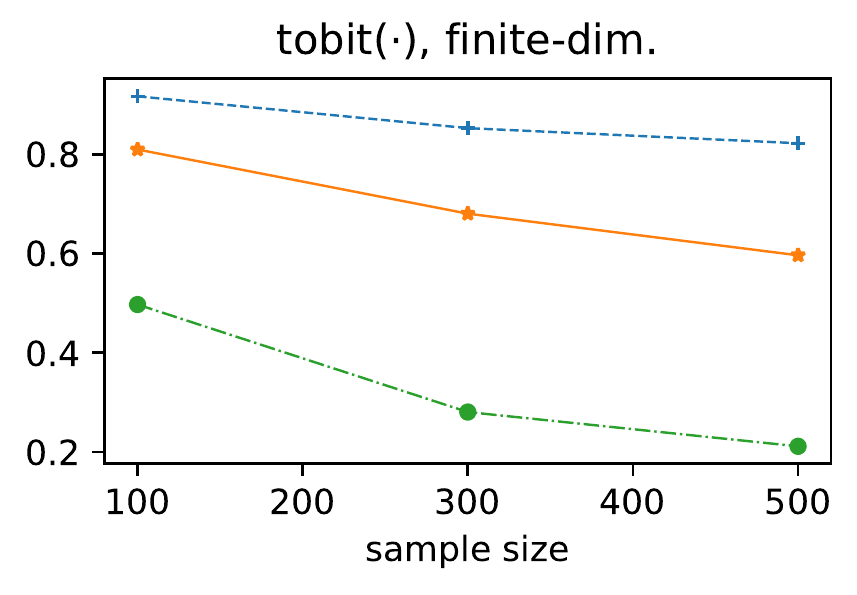}&
				\includegraphics[width=0.36\textwidth,height=0.19\textheight]{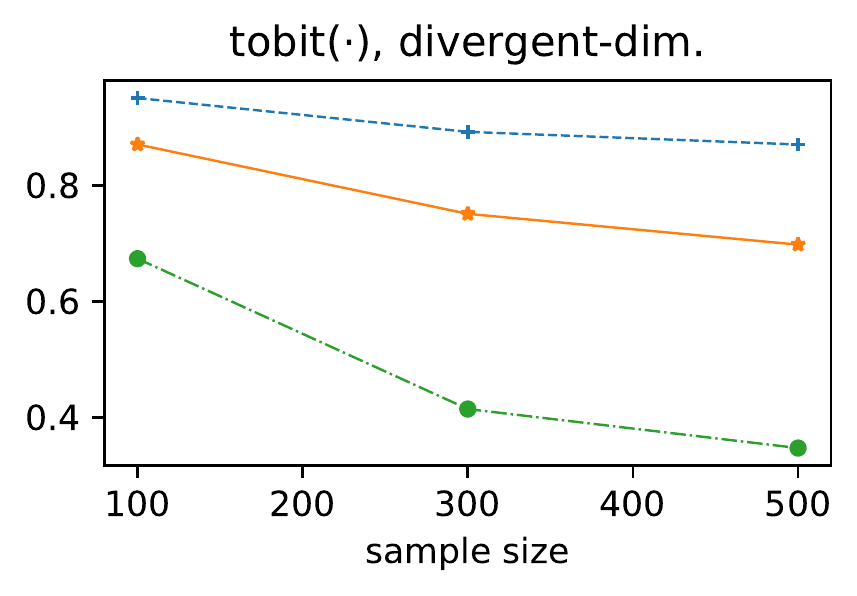}
			\end{tabular}
		\end{center}
		\footnotesize{\emph{Note:} The upper four figures plot the sum of weights assigned to correct models for JCVMA, namely, $w_{\Delta}$. The bottom four figures plot the relative squared loss of JCVMA with respect to the best possible averaging estimators using only misspecified models, namely, $L_n(\hat{\w})/\inf_{\w\in \calW_F}L_n(\w)$.}
	\end{figure}

	When the correct models are included in the candidate model set, Theorem~\ref{thm:consist} shows that the weights of JCVMA assigned to the correct models asymptotically tend to one. We verify this theorem by plotting the sum of weights assigned to the correct models against the sample size in the upper panel of Figure~\ref{fig:weight&loss-correct}. Generally, we find that the sum of these weights is monotonically increasing and converges to one as $n$ enlarges. When we increase $R^2$, both the sum of weights on correct models and the speed of convergence of this sum improve. These results confirm the validity of Theorem~\ref{thm:consist}.
	
	An important implication of the weight consistency is that JCVMA has smaller squared loss than other estimators that average the misspecified models as shown in Corollary~\ref{thm:opt-correctmodel}. The bottom four figures of Figure~\ref{fig:weight&loss-correct} verify this corollary by plotting the relative squared loss with respect to the best possible averaging estimators that only use misspecified models. The relative squared loss of JCVMA is indeed less than one and generally decreases as $n$ increases. We also see that the relative squared loss is lower when $R^2$ is larger.

	\subsection{Simulation under $p>n$}
	\label{sec:sub54}
	
	Thus far, we have studied the finite-sample performance of JCVMA when $p<n$. Now we consider 
	situations in which $p$ is larger than $n$. We set $n=100$ and $p=200$. The coefficient vector is set as $\bbeta=(1,2, 0.1,3,0.08,4,0.06,5,0.04,6,0.02,0,0,0,0,0,\ldots,0,4)\tt
	$, which is characterized by sparsity. We consider the misspecified scenario, where the last covariate is omitted by all fitted models. 
	
	Since the full model contains more parameters than the number of observations, it cannot be estimated by standard NLS.
	An excessively large $p$ also implies that there are a formidable number of candidate models if we consider all possible combinations of covariates.
	Hence, we employ the regularization method with an $L_1$ penalty to estimate the full SIM, and pre-screen candidate models using the second approach (regularization-based screening) discussed in Section~\ref{sec:pbigern}. Particularly, we vary the tuning parameter $\lambda$ by taking 10 evenly spaced points between 0.001 (which yields on average 150 non-zero coefficient estimates across replications) and 0.02 (which forces all coefficient estimates to be zero across replications). Such a variation of $\lambda$ leads to 10 candidate models, over which all selection and averaging methods are applied to predict the response variable.
	
\begin{figure}[H]
		\begin{center}		\caption{NMSPE when $p>n$ and all candidate models are misspecified}
			\label{fig:pbiggern}
			\begin{tabular}{cc} \includegraphics[scale=0.75]{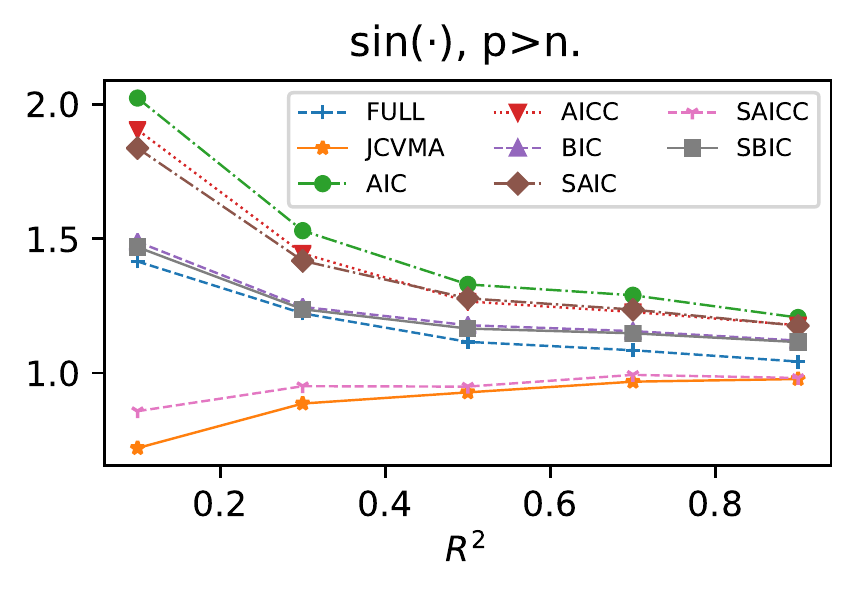}&		\includegraphics[scale=0.75]{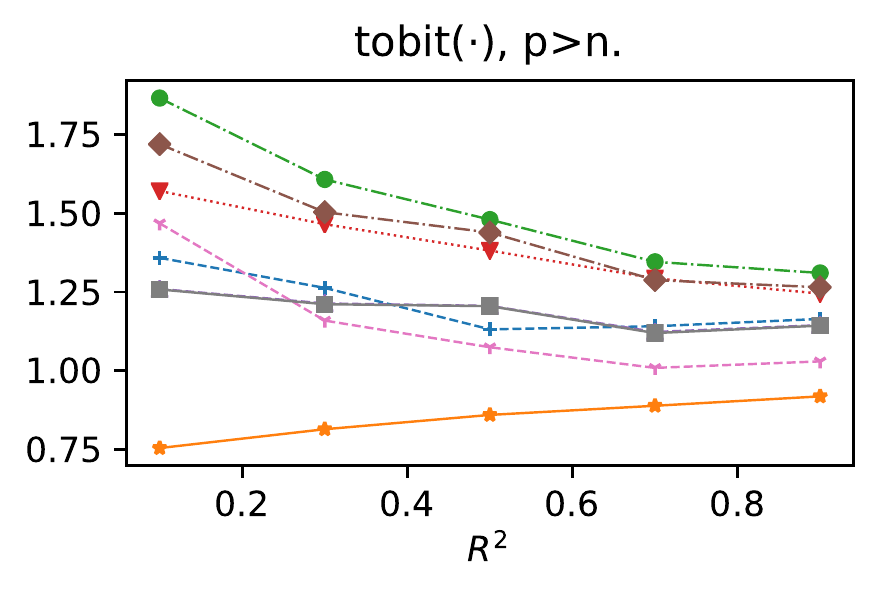}
			\end{tabular}
		\end{center}
		\scriptsize{\emph{Note:} We set $p=200$ and $n=100$. The NMSPE is defined as $D^{-1}\sum_{d=1}^{D}L_n^{(d)}/L^{(d)}_{\min}$. }
	\end{figure}

	Figure~\ref{fig:pbiggern} presents the NMSPEs of the competing methods when $p>n$ and all candidate models are misspecified. It shows that JCVMA based on regularized estimation and pre-screening outperforms other selection and averaging methods, and again its advantage is particularly prominent when $R^2$ is small. When $R^2$ is large, the squared losses of all methods are asymptotically identical to that of the best single model because all of the candidate models perform similarly after this pre-screening.
	
	\section{Empirical application: Financial development and income distribution}\label{sec:empirical}

	Given the substantial cross-country difference in inequality and the level of financial development, it is of particular interest for both academics and policy makers to understand whether and how financial development affects the income distribution.
	In this section, we revisit the relationship between financial development and the distribution of income, first studied by \citet{beck2007}. Our response variable is the growth rate of Gini coefficient ($G$). 
	We measure financial development by private credit ($P$), which is a logarithm of credit by financial intermediaries to the private sector divided by GDP. Other explanatory variables include the logarithm of the initial Gini coefficient ($G_{init}$), initial human capital stock ($H_{init}$) measured by the logarithm of secondary school attainment in the initial year, international openness ($O$) measured by the sum of exports and imports divided by GDP, and inflation ($I$). See \citet{beck2007} for more details on the variable definitions and constructions. We employ the same dataset as \citet{beck2007}, which covers 78 countries over the period from 1958 to 1997. After deleting missing values, we obtain a sample containing $n=256$ observations.
	
	Economic theory suggests that the impact of financial development on income distribution may be two-fold. First, improvement in the financial system is expected to reduce inequality, because financial imperfections impose particularly great constraints for the poor who lack collateral and credit histories, and any relaxation of these constraints would disproportionately benefit the poor \citep{beck2010}. In contrast, there are also arguments that financial development primarily benefits the rich, while the poor are less affected by the improvement in the formal financial sectors because they mostly rely on informal financial sources, e.g., family connections.
	Due to such possible two-fold effects, the (net) impact of financial development on the income distribution is likely to be highly nonlinear, as suggested by \citet{greenwood.jovanovic:1990}.

	{To model the potentially nonlinear relation between financial development and income distribution and account for the model uncertainty, we apply the proposed JCVMA to the SIM with two sets of covariates. The first includes the five covariates $(P,G_{init},H_{init},O,I)$ in \citet{beck2007}, leading to $2^5-1=31$ candidate models, and the second additionally includes the multiplicative terms of every two covariates to control for potential interaction effects, thus containing 15 regressors in total and leading to $2^{15}-1=32767$ models if one considers all possible combinations of 15 regressors. In the second case, estimating and averaging all possible models is computationally formidable, and thus we employ the ordering-based pre-screening
		discussed
		in Section~\ref{sec:pbigern}. To estimate candidate SIMs, we use the same choice of $M_n=50$ and cross-validated optimal bandwidth as in the simulation, but adapt the searching range to the real data.}
	
\begin{figure}[ht]
		\begin{center}		\caption{Growth of Gini coefficient: True vs. estimated values}
			\label{fig:nonlinear}\vspace*{-0.2cm}
			\begin{tabular}{ll}
				\multicolumn{1}{c}{\quad The case of 5 covariates}&			\multicolumn{1}{c}{\quad The case of 15 covariates}\\
				\includegraphics[scale=0.25]{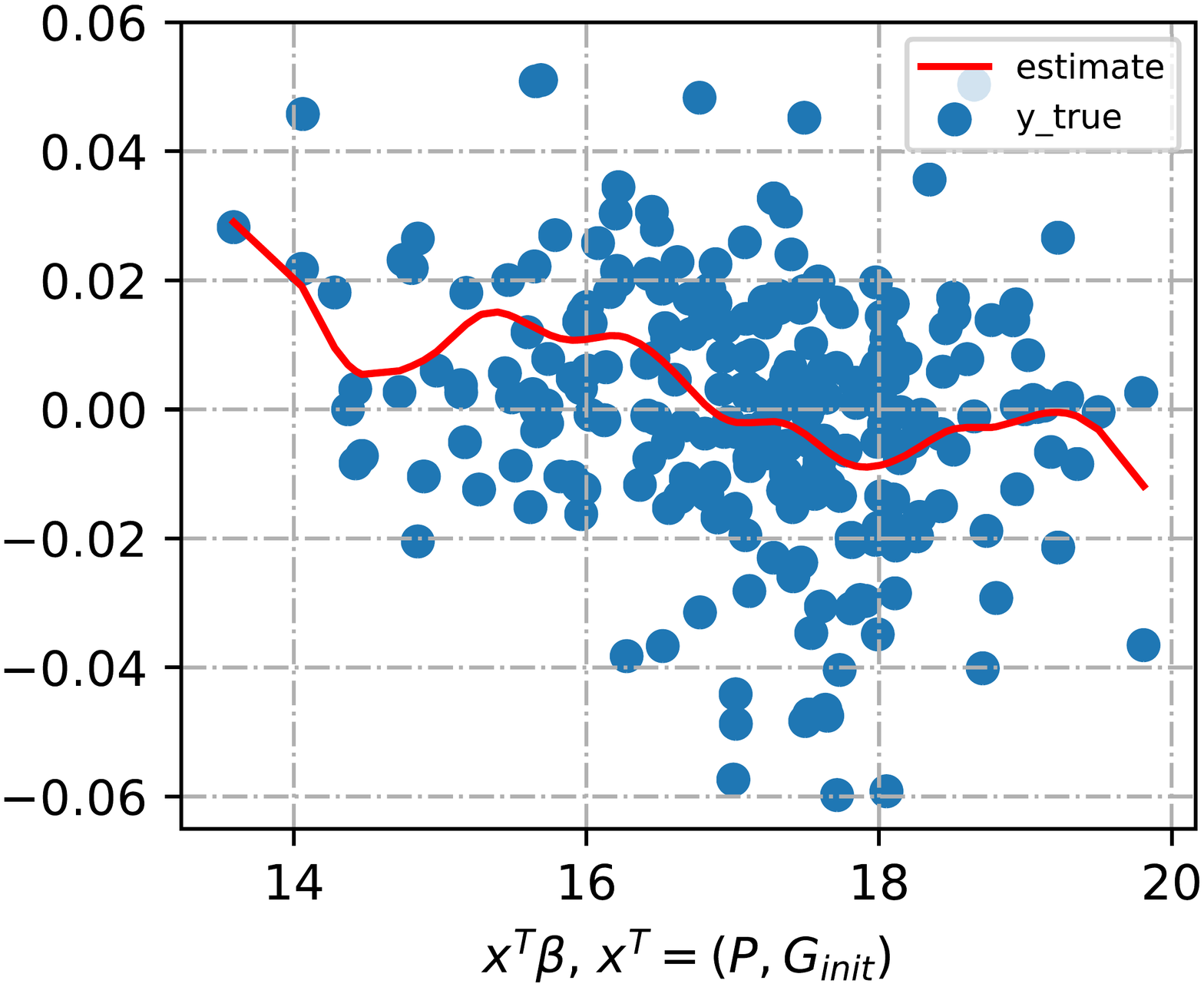}&
				
				\includegraphics[scale=0.25]{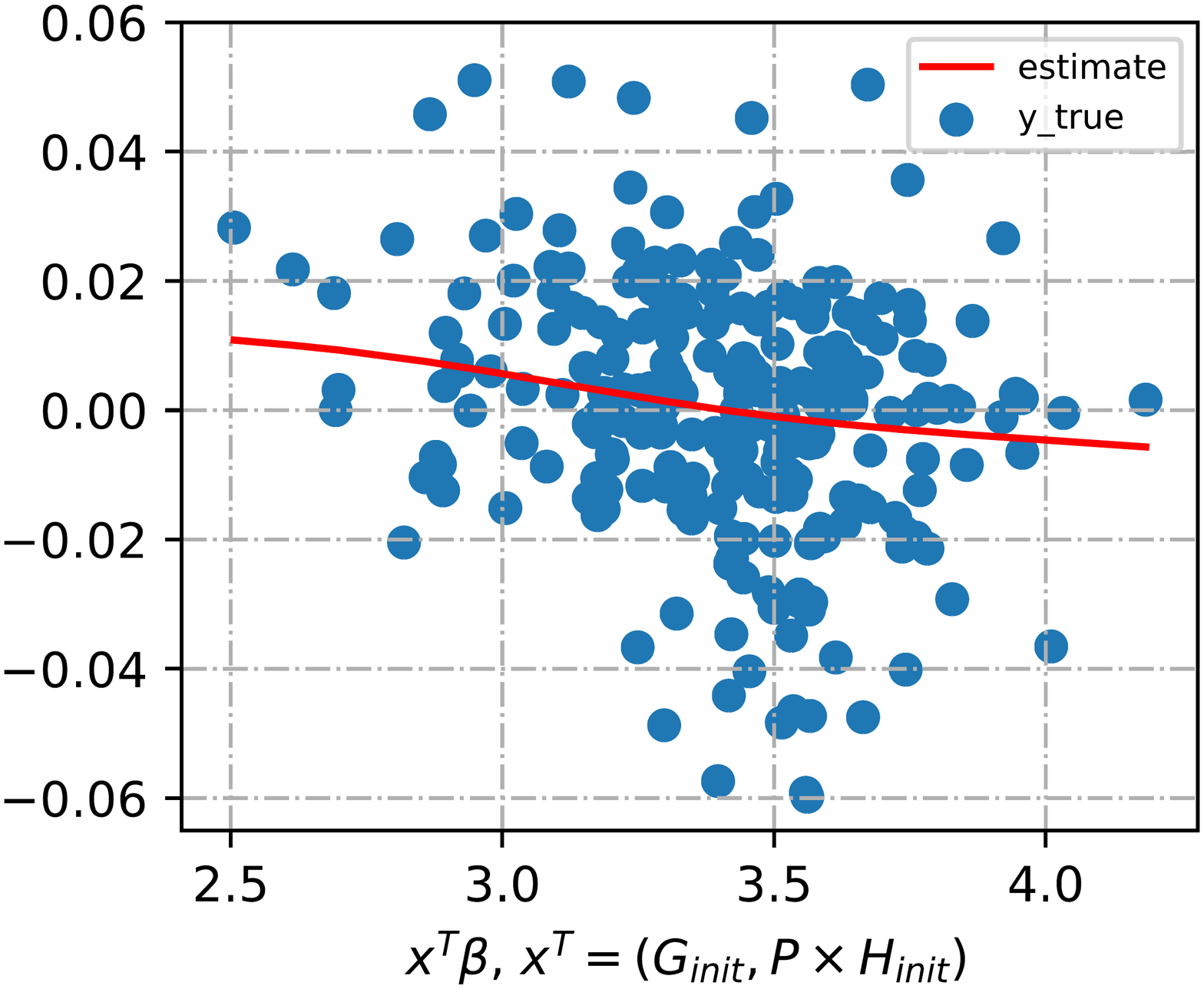}
			\end{tabular}
		\end{center}
		\footnotesize{\emph{Note:} This figure plots the true and predicted values of the Gini coefficient growth against the linear function $\x^T\bbeta$ for the two most weighted candidate SIMs.}
	\end{figure}

	We first examine the effect of financial development on the income distribution. We observe that, for both sets of covariates, most weights ($>0.9$) produced by JCVMA concentrate on only two candidate models, one of which contains financial development ($P$), the variable of interest. The total effect of a covariate in SIMs should be jointly inferred by the coefficient estimates and the estimated link function. We find that for the model that includes financial development, the estimated coefficient of financial development is significantly positive at the 5\% level (subject to normalization), where the confidence interval based on JCVMA is obtained by bootstrapping with 500 resamplings. Figure~\ref{fig:nonlinear} plots the true and predicted values of the Gini coefficient growth against the linear function $\x^T\bbeta$ for the most heavily weighted candidate SIMs, which includes financial development.
	It is revealed that the estimated link function is positive for small values of $\x^T\bbeta$ but negative when $\x^T\bbeta$ is moderate or large.
	%
	%
	These estimation results jointly imply that for a portion of observations, the effect of financial development on the growth of Gini coefficient is significantly negative, which explains the negative overall effect of OLS as reported by \citet{beck2007}. However, this effect is significantly positive for observations with relatively small values of $\x^T\bbeta$. Further examination suggests that observations with positive effects of financial development are typically characterized by much higher growth rates of Gini coefficient and inflation than those with negative effects. The variability of the link function implies that financial development does help alleviate income inequality when the degree of inequality is stable with little inflation, but in some countries, e.g., Korea, Indonesia, and several European countries in the 1960s-1970s with particularly high inflation, financial development further accelerates the growth of inequality. Our results are consistent with the economic theory that financial development exerts two-fold effects depending on the economic and social status \citep{beck2010,greenwood.jovanovic:1990}.
	
	Now, we examine the performance of JCVMA in predicting the growth of Gini coefficient. 
	We consider the pseudo out-of-sample prediction over time. We divide the entire time period into two subsamples at 1987, 1988, 1989, 1990, 1991, and 1993, so that the first subsample, which is used as the training set, consists of approximately 60\%, 65\%, 70\%, 75\%, 80\% and 85\% of the entire sample, respectively.
	Accordingly, the second subsample is used as the testing set.
	We estimate the parameters, link functions and weights of each candidate model from the training set, and then use them to predict the response variable in the testing set. We follow \citet{Hansen2008Least} to evaluate the competing methods according to the mean squared prediction error (MSPE) defined as
	$
	\textrm{MSPE}=n_{\textrm{test}}^{-1} \left\|\hat{\y}-\y_{\textrm{test}} \right\|^2-\widehat{\sigma}^2,
	$
	where $\y_{\textrm{test}}$ is the response variable of the testing set, $\hat{\y}$ is its predicted value, and $\widehat{\sigma}^2=(n-1)^{-1}\sumin(y_i-\bar{y})^2$ is the estimated variance of $y_i$ based on the entire sample with $\bar{y}$ being the sample mean of $y_i$. The results are presented in Table~\ref{tab:SRMSPE-realdata}. All numbers are divided by the MSPE of the full SIM, so that a value smaller than 1 suggests a better prediction than the full SIM. Table \ref{tab:SRMSPE-realdata} shows that JCVMA performs the best in {9 of 12 cases}. Even when JCVMA does not produce the lowest MSPE, it is close to the best method, suggesting its robustness, while the performances of other methods vary across cases to a large extent.
	
	\begin{table}[htp]
		\begin{center}
			\caption{MSPE of growth of Gini coefficient}\footnotesize
			\label{tab:SRMSPE-realdata}%
			\begin{tabular}{lcccccccccccccc}
				\hline\hline
				\makecell{Training\\ sample size}  & \makecell{$\approx 0.6n$\\154}   & \makecell{$\approx 0.65n$\\165}  & \makecell{$\approx 0.7n $\\173}  & \makecell{$\approx 0.75n$\\190}   &\makecell{$\approx 0.8n $\\197} &\makecell{$\approx 0.85n $\\216}\\
				\hline
				& \multicolumn{6}{c}{5 covariates}\\
				JCVMA & 0.860  & \textbf{0.643} & \textbf{0.904}   & \textbf{0.893}  & \textbf{0.907} & 0.901               \\			
				AIC   & 0.849           & 0.646          & 1.435            & 1.179           & 0.924          & \textbf{0.794}      \\
				BIC   & \textbf{0.840}           & 0.646          & 0.911            & 0.909           & 0.912          & 0.995               \\
				AICC  & \textbf{0.840}           & 0.646          & 0.911            & 0.904           & 0.924          & \textbf{0.794}      \\
				SAIC  & 0.865           & 0.644          & 1.263            & 1.010           & 0.918          & \textbf{0.794}      \\
				SBIC  & \textbf{0.840}           & 0.646          & 0.911            & 0.909           & 0.921          & 0.993               \\
				SAICC & 0.871           & 0.692          & 0.951            & 0.953           & 0.982          & 1.066               \\
				Full  & 1.000           & 1.000          & 1.000            & 1.000           & 1.000          & 1.000               \\
				\hline
				& \multicolumn{6}{c}{15 covariates}\\
				JCVMA & \textbf{0.983}  & \textbf{0.748} & 0.892            & \textbf{0.864}  & \textbf{0.616} & \textbf{0.559}      \\			
				AIC   & 1.000           & 0.772          & 0.840            & 1.045           & 0.888          & 0.915      \\
				BIC   & 0.997           & 0.752          & \textbf{0.763}            & 0.901           & 0.644          & 0.560      \\
				AICC  & 1.676           & 1.106          & 0.840            & 1.745           & 0.888          & 0.560      \\
				SAIC  & 1.000           & 0.772          & 0.840            & 1.045           & 0.888          & 0.915      \\
				SBIC  & 0.997           & 0.752          & \textbf{0.763}            & 0.901           & 0.644          & 0.560      \\
				SAICC & 1.676           & 1.106          & 0.840   & 1.745           & 0.781          & 0.594      \\
				Full  & 1.000           & 1.000          & 1.000            & 1.000           & 1.000          & 1.000      \\
				
				\hline
			\end{tabular}%
		\end{center}
		\scriptsize{\emph{Note:} 
		The value in bold indicates the minimum in a column. The upper panel considers 5 covariates, namely, ($P$, $G_{init}$, $H_{init}$, $O$, and $I$), and the bottom panel considers 15 covariates including both level and interaction terms.}
	\end{table}%

	To summarize, our empirical application shows that the proposed JCVMA, which accounts for nonlinearity and model uncertainty, provides a useful tool for prediction and produces new economic insights on the relationship between financial development and the income distribution.

	\section{Concluding remarks}\label{sec:conclusion}
	This paper proposes a model-averaging method to address the model uncertainty in single-index models, and our averaging method allows the numbers of covariates and candidate models to diverge when the sample size increases. {We also propose model averaging based on regularized estimation and pre-screening to deal with many covariates and candidate models.} We demonstrate the superior properties of the proposed method when all candidate models are misspecified and when correct models are available in the candidate model set.
	
	{Our proposed approach averages the predicted values of response variables. An alternative way of constructing averaging estimators for SIM is to average estimators of $\bbeta$. Such an approach provides a good interpretation of slope coefficients, while it is less flexible than the approach of this paper since it imposes a common link function $g(\cdot)$ for all candidate models. In contrast, averaging predicted values allows the link functions $ g_\s(\cdot)$ to vary across candidate models, and this flexibility often helps improve the prediction accuracy. Given that the goal of optimal model averaging is typically prediction, we adopt our current averaging strategy, but averaging coefficient estimators can be useful in other contexts and deserves further study. Another possible direction for future research is to extend the current averaging technique to other varieties of SIMs, such as multiple-index models with different numbers of indices, single-index varying-coefficient models and generalized partially linear SIMs.
	Finally, how to properly choose $J_n$ or $M_n$ in our cross-validation procedure also warrants future work.}


	\bibliographystyle{abbrvnat}
	{
		\bibliography{MA_SIM_ref2}

\begin{thebibliography}{41}
\providecommand{\natexlab}[1]{#1}
\providecommand{\url}[1]{\texttt{#1}}
\expandafter\ifx\csname urlstyle\endcsname\relax
  \providecommand{\doi}[1]{doi: #1}\else
  \providecommand{\doi}{doi: \begingroup \urlstyle{rm}\Url}\fi

\bibitem[Ando and Li(2014)]{Ando2014}
T.~Ando and K.-C. Li.
\newblock A model-averaging approach for high-dimensional regression.
\newblock \emph{Journal of the American Statistical Association}, 109:\penalty0
  254--265, 2014.

\bibitem[Ando and Li(2017)]{Ando2017}
T.~Ando and K.-C. Li.
\newblock A weight-relaxed model averaging approach for high-dimensional
  generalized linear models.
\newblock \emph{The Annals of Statistics}, 45:\penalty0 2654--2679, 2017.

\bibitem[Beck et~al.(2007)Beck, Demirg\"{u}\c{c}-Kunt, and Levine]{beck2007}
T.~Beck, A.~Demirg\"{u}\c{c}-Kunt, and R.~Levine.
\newblock Finance, inequality and the poor.
\newblock \emph{Journal of Economic Growth}, 12:\penalty0 27--49, 2007.

\bibitem[Beck et~al.(2010)Beck, Levine, and Levkov]{beck2010}
T.~Beck, R.~Levine, and A.~Levkov.
\newblock Big bad banks? {T}he winners and losers from bank deregulation in the
  {U}nited {S}tates.
\newblock \emph{The Journal of Finance}, 65:\penalty0 1637--1667, 2010.

\bibitem[Buckland et~al.(1997)Buckland, Burnham, and Augustin]{Buckland:1997}
S.~T. Buckland, K.~P. Burnham, and N.~H. Augustin.
\newblock Model selection: An integral part of inference.
\newblock \emph{Biometrics}, 53:\penalty0 603--618, 1997.

\bibitem[Cheng et~al.(2017)Cheng, Zeng, and Zhu]{cheng.zeng:2017}
L.~Cheng, P.~Zeng, and Y.~Zhu.
\newblock {BS-SIM}: An effective variable selection method for high-dimensional
  single index model.
\newblock \emph{Electronic Journal of Statistics}, 11:\penalty0 3522--3548,
  2017.

\bibitem[Claeskens et~al.(2006)Claeskens, Croux, and {van
  Kerckhoven}]{claeskens.croux.ea:2006}
G.~Claeskens, C.~Croux, and J.~{van Kerckhoven}.
\newblock Variable selection for logistic regression using a prediction-focused
  information criterion.
\newblock \emph{Biometrics}, 62:\penalty0 972--979, 2006.

\bibitem[Fan and Gijbels(1996)]{Fan1996Local}
J.~Fan and I.~Gijbels.
\newblock \emph{Local Polynomial Modelling and Its Applications}.
\newblock Chapman \& Hall, London, 1996.

\bibitem[Fan and Lv(2008)]{fan&lv:2008}
J.~Fan and J.~Lv.
\newblock Sure independence screening for ultrahigh dimensional feature space.
\newblock \emph{Journal of the Royal Statistical Society: Series B},
  70:\penalty0 849--911, 2008.

\bibitem[Friedman et~al.(2010)Friedman, Hastie, and
  Tibshirani]{Friedman2010Regularization}
J.~H. Friedman, T.~Hastie, and R.~Tibshirani.
\newblock Regularization paths for generalized linear models via coordinate
  descent.
\newblock \emph{Journal of Statistical Software}, 33:\penalty0 1--22, 2010.

\bibitem[Greenwood and Jovanovic(1990)]{greenwood.jovanovic:1990}
J.~Greenwood and B.~Jovanovic.
\newblock Financial development, growth and the distribution of income.
\newblock \emph{Journal of Political Economy}, 98:\penalty0 1076--1107, 1990.

\bibitem[Hansen(2007)]{hansen2007least}
B.~E. Hansen.
\newblock Least squares model averaging.
\newblock \emph{Econometrica}, 75:\penalty0 1175--1189, 2007.

\bibitem[Hansen(2008)]{Hansen2008Least}
B.~E. Hansen.
\newblock Least-squares forecast averaging.
\newblock \emph{Journal of Econometrics}, 146:\penalty0 342--350, 2008.

\bibitem[Hansen(2014{\natexlab{a}})]{hansen2014book}
B.~E. Hansen.
\newblock Nonparametric sieve regression, least squares, averaging least
  squares and cross-validation.
\newblock In J.~S. Racine, L.~Su, A.~Ullah, and B.~E. Hansen, editors,
  \emph{The Oxford Handbook of Applied Nonparametric and Semiparametric
  Econometrics and Statistics}. Oxford University Press, 2014{\natexlab{a}}.

\bibitem[Hansen(2014{\natexlab{b}})]{hansen:2014}
B.~E. Hansen.
\newblock Model averaging, asymptotic risk and regressor groups.
\newblock \emph{Quantitative Economics}, 5:\penalty0 495--530,
  2014{\natexlab{b}}.

\bibitem[Hansen and Racine(2012)]{hansen.racine:2012}
B.~E. Hansen and J.~Racine.
\newblock Jackknife model averaging.
\newblock \emph{Journal of Econometrics}, 167:\penalty0 38--46, 2012.

\bibitem[H\"ardle et~al.(2007)H\"ardle, Liang, and Gao]{Hardle2000Partially}
W.~K. H\"ardle, H.~Liang, and J.~Gao.
\newblock \emph{Partially linear models}.
\newblock Springer, Berlin, Heidelberg, 2007.

\bibitem[Hoeting et~al.(1999)Hoeting, Madigan, Raftery, and
  Volinsky]{hoeting.madigan.ea:1999}
J.~A. Hoeting, D.~Madigan, A.~E. Raftery, and C.~T. Volinsky.
\newblock Bayesian model averaging: {A} tutorial.
\newblock \emph{Statistical Science}, 14:\penalty0 382--417, 1999.

\bibitem[Horowitz(1998)]{Horowitz1998Semiparametric}
J.~L. Horowitz.
\newblock \emph{Semiparametric Methods in Econometrics}.
\newblock Springer, 1998.

\bibitem[Ichimura(1993)]{Ichimura1993Semiparametric}
H.~Ichimura.
\newblock Semiparametric least squares ({SLS}) and weighted {SLS} estimation of
  single-index models.
\newblock \emph{Journal of Econometrics}, 58:\penalty0 71--120, 1993.

\bibitem[Kong and Xia(2007)]{kong.xia:2007}
E.~Kong and Y.~Xia.
\newblock Variable selection for the single-index model.
\newblock \emph{Biometrika}, 94:\penalty0 217--229, 2007.

\bibitem[Leng(2010)]{2010Variable}
C.~Leng.
\newblock Variable selection and coefficient estimation via regularized rank
  regression.
\newblock \emph{Statistica Sinica}, 20:\penalty0 167--181, 2010.

\bibitem[Li et~al.(2018)Li, Li, Racine, and Zhang]{li.li.racine.zhang:2018}
C.~Li, Q.~Li, J.~S. Racine, and D.~Zhang.
\newblock Optimal model averaging of varying coefficient models.
\newblock \emph{Statistica Sinica}, 28:\penalty0 2795--2809, 2018.

\bibitem[Liang et~al.(2007)Liang, Wang, and Carroll]{liang.wang.carroll:2007}
H.~Liang, S.~Wang, and R.~J. Carroll.
\newblock Partially linear models with missing response variables and
  error-prone covariates.
\newblock \emph{Biometrika}, 94:\penalty0 185--198, 2007.

\bibitem[Liang et~al.(2010)Liang, Liu, Li, and Tsai]{LiangEstimation}
H.~Liang, X.~Liu, R.~Li, and C.-L. Tsai.
\newblock Estimation and testing for partially linear single-index models.
\newblock \emph{The Annals of Statistics}, 38:\penalty0 3811--3836, 2010.

\bibitem[Liu(2015)]{LiuDistribution}
C.-A. Liu.
\newblock Distribution theory of the least squares averaging estimator.
\newblock \emph{Journal of Econometrics}, 186:\penalty0 142--159, 2015.

\bibitem[Liu(2018)]{liu:2018}
C.-A. Liu.
\newblock Averaging estimators for kernel regressions.
\newblock \emph{Economics Letters}, 171:\penalty0 102--105, 2018.

\bibitem[Liu et~al.(2021)Liu, Zou, Zhao, and Yang]{Liu2021}
Y.~Liu, J.~Zou, S.~Zhao, and Q.~Yang.
\newblock Model averaging estimation for varying-coefficient single-index
  models.
\newblock \emph{Journal of Systems Science and Complexity}, forthcoming, 2021.

\bibitem[Naik and Tsai(2001)]{naik:tsai:2001}
P.~A. Naik and C.-L. Tsai.
\newblock Single-index model selections.
\newblock \emph{Biometrika}, 88:\penalty0 821--832, 2001.

\bibitem[Powell et~al.(1989)Powell, Stock, and Stoker]{PowellSemiparametric}
J.~L. Powell, J.~H. Stock, and T.~M. Stoker.
\newblock Semiparametric estimation of index coefficients.
\newblock \emph{Econometrica}, 57:\penalty0 1403--1430, 1989.

\bibitem[Radchenko(2015)]{Radchenko2015High}
P.~Radchenko.
\newblock High dimensional single index models.
\newblock \emph{Journal of Multivariate Analysis}, 139:\penalty0 266--282,
  2015.

\bibitem[Wan et~al.(2010)Wan, Zhang, and Zou]{wan2010least}
A.~T.~K. Wan, X.~Zhang, and G.~Zou.
\newblock Least squares model averaging by {M}allows criterion.
\newblock \emph{Journal of Econometrics}, 156:\penalty0 277--283, 2010.

\bibitem[Wang et~al.(2011)Wang, Xue, Zhu, and Chong]{Wang2010}
J.~Wang, L.~Xue, L.~Zhu, and Y.~S. Chong.
\newblock Estimation for a partial-linear single-index model.
\newblock \emph{The Annals of Statistics}, 38:\penalty0 246--274, 2011.

\bibitem[White(1982)]{White1982}
H.~White.
\newblock Maximum likelihood estimation of misspecified models.
\newblock \emph{Econometrica}, 50:\penalty0 1--25, 1982.

\bibitem[Yu et~al.(2014)Yu, He, and Chen]{YuEmpirical}
Z.~Yu, B.~He, and M.~Chen.
\newblock Empirical likelihood for generalized partially linear single-index
  models.
\newblock \emph{Communication in Statistics-Theory and Methods}, 43:\penalty0
  4156--4163, 2014.

\bibitem[Yuan and Yang(2005)]{yuan.yang:2005}
Z.~Yuan and Y.~Yang.
\newblock Combining linear regression models: When and how?
\newblock \emph{Journal of the American Statistical Association}, 100:\penalty0
  1202--1214, 2005.

\bibitem[Zhang and Wang(2019)]{zhang.wang:2019}
X.~Zhang and W.~Wang.
\newblock Optimal model averaging estimation for partially linear models.
\newblock \emph{Statistica Sinica}, 29:\penalty0 693--718, 2019.

\bibitem[Zhang et~al.(2016)Zhang, Yu, Zou, and Liang]{Xinyu2016Optimal}
X.~Zhang, D.~Yu, G.~Zou, and H.~Liang.
\newblock Optimal model averaging estimation for generalized linear models and
  generalized linear mixed-effects models.
\newblock \emph{Journal of the American Statistical Association}, 111:\penalty0
  1775--1790, 2016.

\bibitem[Zhang et~al.(2020)Zhang, Zou, Liang, and
  Carroll]{zhang2019Parsimonious}
X.~Zhang, G.~Zou, H.~Liang, and R.~J. Carroll.
\newblock Parsimonious model averaging with a diverging number of parameters.
\newblock \emph{Journal of the American Statistical Association}, 115:\penalty0
  972--984, 2020.

\bibitem[Zhu et~al.(2019)Zhu, Wan, Zhang, and Zou]{ZhuA}
R.~Zhu, A.~T.~K. Wan, X.~Zhang, and G.~Zou.
\newblock A {M}allows-type model averaging estimator for the
  varying-coefficient partially linear model.
\newblock \emph{Journal of the American Statistical Association}, 114:\penalty0
  882--892, 2019.

\bibitem[Zou(2006)]{Zou2006The}
H.~Zou.
\newblock The adaptive lasso and its oracle properties.
\newblock \emph{Journal of the American Statistical Association}, 101:\penalty0
  1418--1429, 2006.

\end{thebibliography}
	}
	%

	\clearpage
	\section*{Appendix}
	\setcounter{equation}{0}
	\setcounter{page}{1}
	\setcounter{subsection}{0}
	\renewcommand{\theequation}{A.\arabic{equation}}
	\renewcommand{\thesubsection}{A.\arabic{subsection}}
	\baselineskip=18pt

	\renewcommand{\theasu}{A\arabic{subsection}.\arabic{asu}}
	\setcounter{asu}{0}

	This appendix provides additional conditions for Lemma~\ref{lem:consistance} and Corollary~\ref{thm:high} in the paper.
	
	\subsection{Conditions for Lemma~\ref{lem:consistance}}
	\label{sec:conditions}
	
	The following regularity conditions are required for the consistency of the NLS estimator and its cross-validation version for each candidate model. 

	\begin{asu}\label{con:conlast}
		\subasu\label{eq:last0}
		The kernel function $k(s)$ is a bounded symmetric density with a compact support; \subasu\label{eq:last2} The following quantities are finite: $\int |\tau^3| k(\tau)\md \tau$, $\int |\tau k'(\tau)|\md \tau$, $\int \tau^2 |k'(\tau)|\md \tau$, $\int k^{'2}(\tau)\md \tau$, $\int |\tau| k^{'2}(\tau)\md \tau$ and $\int \tau^2 k^{'2}(\tau)\md \tau$, where $k^{'}(s)$ is the first-order derivative of $k(s)$.
	\end{asu}
	These are common restrictions on the kernel function in nonparametric statistics, such as Lemmas .2--.4 in \cite{Ichimura1993Semiparametric} and Condition (C.5) in \cite{ZhuA}.

	\begin{asu}\label{con:hs}\
		\subasu\label{eq:hs1} $\sups h_s\to 0$;
		\subasu\label{eq:hs2} $\sumsn n^{-1} h_s^{-3}p_s=O_P(1)$;\\
		\subasu\label{eq:hs3} $\infs n h_s\to\infty$; {\subasu\label{eq:hs4} $\sups (nh_s^4+h_s^{-1})/M_n^2 n d_n^2=O(1)$.}
	\end{asu}
	This condition pertains to the bandwidth of the averaging estimator. $S_n$ and $p_s$ appear because we need to solve $S_n$ candidate models simultaneously. The similar conditions are also used in Condition (C5) in \cite{Wang2010} and Condition (C.5) in \cite{ZhuA}.
	

	\begin{asu}\label{con:mu-gs}\
		\begin{itemize}
			\item[]\subasu\label{eq:gs1}
			There exists a universal constant $\bar{C}>0$ such that  $\sups\supi\|\x_{\s,i}\|\leq \sqrt{p_s} \bar{C}$;
			\item[]\subasu\label{eq:gs22}
			$\sups\supi|\x_{\s,i}\tt\sbeta_\s|<\infty$ and $\sups\supi|x_{\s,ir}\beta^*_r|<\infty$;
			\item[]\subasu\label{eq:gs2}
			$\sups\supi\left|g_\s(\x_{\s,i}\tt\sbeta_\s) \right|= O_P(1)$;
			\item[]\subasu\label{eq:gs3}
			There exists a constant $\underline{c}$ such that $\infs \min_{r:1\leq r\leq p_s, \beta^*_{\s,r}\neq 0} |\beta^*_{\s,r}|>\underline{c}>0$;
			\item[]\subasu\label{eq:partial}
			$\sups\frac{1}{\sqrt{nS_np_s}}\left\|
			\frac{\partial}{\partial \bbeta_\s}
			\sumin \left\{\mu_i-\frac{1}{n-1}\sum_{j\neq i}^nK_{\s,ij}(\sbeta_\s)\mu_j\right\}^2
			\right\|=O_P(1).$
		\end{itemize}
	\end{asu}
	Condition \ref{eq:gs1} holds if each element of $\x_i$ is uniformly bounded, an assumption also imposed by \citeauthor{Radchenko2015High} (2015, Assumption~A1). Conditions \ref{eq:gs22} and \ref{eq:gs2} require that the quasi-true parameter is not abnormal so that the estimator for $\bbeta_\s$ is {well-behaved}.
	Condition \ref{eq:gs3} guarantees that the nonzero parameters, $\beta^*_{\s,r}$, have a uniform lower bound. 
	Condition \ref{eq:partial} requires that the difference between $\bmu$ and the theoretical estimator from the $s\th$ candidate model, $\K_\s(\sbeta_\s)\bmu$, is smooth enough around $\sbeta_\s$ such that there is sufficient information to estimate the quasi-true parameter $\sbeta_\s$. {$\sqrt{p_s}$ and $\sqrt{S_n}$ appear in the left-side denominator in this condition, since $\left\|
		\partial\sumin \left\{\mu_i-\frac{1}{n-1}\sum_{j\neq i}^nK_{\s,ij}(\sbeta_\s)\mu_j\right\}^2/\partial \bbeta_\s\right\|$ is of order $\sqrt{p_s}$ and there are $S_n$ candidate models}.

	Let $\rho_\s(v_1,v_2,\ldots,v_{p_s})$ denote the joint density function of $x_{\s,1}\beta^*_1,x_{\s,2}\beta^*_2,\ldots,x_{\s,p_s}\beta^*_{p_s}$ for the $s\th$ candidate model, where $\x_\s=(x_{\s,1}, x_{\s,2},\ldots,x_{\s,p_s})\tt$ and $\bbeta^*_\s=(\beta^*_1,\beta^*_2,\ldots,\beta^*_{p_s})\tt$.
	Let $f_\s(t)$ denote the density of $\x_{\s}\tt\sbeta_\s$, and denote $f'_\s(t)$ and $f''_\s(t)$ as the first and second-order derivatives of $f_\s(t)$, respectively.
	Further let $\phi_\s(t)=g_\s(t)f_\s(t)$ and $\varphi_\s(t)=g^2_\s(t)f_\s(t)$.
	\begin{asu}\label{con:fs}\
		\begin{itemize}
			\item[] \subasu \label{con:rho}
			There exists a constant $\bar{C}$ such that
			$$\int\rho_{\s}\left(v_{1},\ldots,v_{k-1},t-\sum_{l\neq k}^{p_s}v_{l},v_{k+1},\ldots,v_{p_s}\right)\md v_{1}\ldots\md v_{k-1}\md v_{k+1}\ldots\md v_{p_s}<\bar{C}$$ uniformly for $s$ and $t$;
			\item[] \subasu\label{eq:last6}
			There exist some constants $\underline{c}$ and $\bar{C}$ such that
			$\underline{c}<f_\s(\x_{\s,i}\tt\sbeta_\s) <\bar{C}$ almost surely for $s=1,2,\ldots,S_n$; $i=1,2,\ldots,n$;
			\item[] \subasu\label{eq:last7}
			There exists a universal constant $\bar{C}$ such that $|f'_\s(\x_{\s,i}\tt\sbeta_\s)|<\bar{C}$, $|f''_\s(\x_{\s,i}\tt\sbeta_\s)|<\bar{C}$ almost surely for $s=1,2,\ldots,S_n;$ $i=1,2,\ldots,n$;
			\item[]\subasu\label{con:lip-twice2}
			There exists a constant $G>0$ and $\omega_\s\left(v_1,\ldots,v_{k-1},v_{k+1},\ldots,v_{p_s}\right)>0$ such that
			\begin{align*}
			&\big|\rho_{\s}\left(v_{1},\ldots,v_{k-1},t_1,v_{k+1},\ldots,v_{p_s}\right)
			-
			\rho_{\s}\left(v_{1},\ldots,v_{k-1},t_2,v_{k+1},\ldots,v_{p_s}\right)\big|\\
			&\leq G \omega_\s\left(v_{1},\ldots,v_{k-1},v_{k+1},\ldots,v_{p_s}\right)|t_1-t_2|,
			\end{align*}
			for any $s$ and $k$, where $\int \omega_\s\left(v_{1},\ldots,v_{k-1},v_{k+1},\ldots,v_{p_s}\right)\md v_{1}\ldots\md v_{k-1}\md v_{k+1}\ldots\md v_{p_s}<\infty$ and $\sum_{r=1,r\neq k}^{p_s}\int |v_{l}| \omega_\s\left(v_{1},\ldots,v_{k-1},v_{k+1},\ldots,v_{p_s}\right)\md v_{1}\ldots\md v_{k-1}\md v_{k+1}\ldots\md v_{p_s}<\infty$ uniformly for any $s$;
			\item[] \subasu \label{eq:last52}
			$f'_\s(t)$ and $f''_\s(t)$ satisfy the Lipschitz condition, i.e., there exist two constants $c_1$ and $c_2$ such that
			$|f'_\s(t_1)-f'_\s(t_2)|\leq c_1|t_1-t_2|$ and $|f''_\s(t_1)-f''_\s(t_2)|\leq c_2|t_1-t_2|$;
			\item[] \subasu\label{eq:last8} $\phi'_\s(t)$ and $\varphi'_\s(t)$ satisfy the Lipschitz condition.
		\end{itemize}
	\end{asu}
	
	This condition imposes restrictions on the joint density of
	$x_{\s,1}\beta^*_1,\ x_{\s,2}\beta^*_2,\ldots, \ x_{\s,p_s}\beta^*_{p_s}$ and the density of $\sum_{r=1}^{p_s} x_{\s,r}\beta^*_r$. Especially, when $\{x_{\s,r}\}_{r=1}^{p_s}$ are independent, $f_\s(t)=\int\rho_{\s}(v_{1},\ldots,\\v_{k-1},t-\sum_{l\neq k}^{p_s}v_{l},v_{k+1},\ldots,v_{p_s})\md v_{1}\ldots\md v_{k-1}\md v_{k+1}\ldots\md v_{p_s}$ because of the convolution product.
	To illustrate this condition, we can consider {the simple case where} each $x_{\s,r}$ is i.i.d.~$N(0,1)$, then $\x_{\s,r}\beta^*_r\sim N(0,\beta^{*2}_r)$ and $$f_\s(t)=\int\rho_{\s}\left(v_{1},\ldots,v_{k-1},t-\sum_{l\neq k}^{p_s}v_{l},v_{k+1},\ldots,v_{p_s}\right)\md v_{1}\ldots\md v_{k-1}\md v_{k+1}\ldots\md v_{p_s}$$
	is the density of $N(0,\sum_{r=1}^{p_s}\beta^{*2}_{\s,r})$. In Conditions \ref{con:rho} and \ref{eq:last6}, if $\beta_r^*=1$ for $r=1,2,\ldots,p_s$, then $|f_\s(t)|\leq (2\pi)^{-1/2}$ and we can take $\bar{C}=(2\pi)^{-1/2}$. Condition \ref{eq:last6} is also similar to Condition (C.2) in \cite{ZhuA}.
	
	Condition \ref{eq:last7} ensures that $f'_\s(\x_{\s,i}\tt\sbeta_\s)$ and $f''_\s(\x_{\s,i}\tt\sbeta_\s)$ are both uniformly bounded. From the discussion of Condition \ref{con:rho}, we have $|f'_\s(\x_{\s,i}\tt\sbeta_\s)|=(2\pi e)^{-1/2} \sum_{r=1}^{p_s}(\beta^{*2}_{\s,r})^{-1}\leq (2\pi e)^{-1/2}$ and $|f''_\s(\x_{\s,i}\tt\sbeta_\s)|= (2\pi)^{-1/2}\sum_{r=1}^{p_s}(\beta^{*2}_{\s,r})^{-3/2}\leq (2\pi)^{-1/2}$ uniformly for $s$ and $i$.
	
	%
	%
	Condition \ref{con:lip-twice2} guarantees that the joint density is Lipschitz continuous so that the data are smooth around $x_{\s,r}\tt\beta^{*}_{\s,r}$.
	{For example, if each $x_{\s,r}$ is i.i.d.~$N(0,1)$}, we can derive that
	\begin{align*} &\big|\rho_{\s}(v_{1},\ldots,v_{k-1},t_1,v_{k+1},\ldots,v_{p_s})
	-
	\rho_{\s}\big(v_{1},\ldots,v_{k-1},t_2,v_{k+1},\ldots,v_{p_s}\big)\big|\\
	&\leq \frac{1}{(\sqrt{2\pi})^{p_s-1}\prod_{i=1,i\neq k}^{p_s}\beta_i^*}\exp\left(-\sum_{i=1,i\neq k}^{p_s}\frac{v_i^2}{2\beta_i^{*2}}\right)\times\frac{1}{\sqrt{2\pi}\beta_k^*}
	\bigg|\exp\left(-\frac{t_1^2}{2\beta_k^{*2}}\right)-\exp\left(-\frac{t_2^2}{2\beta_k^{*2}}\right)\bigg|\\
	&\leq \frac{1}{(\sqrt{2\pi})^{p_s-1}\prod_{i=1,i\neq k}^{p_s}\beta_i^*}\exp\left(-\sum_{i=1,i\neq k}^{p_s}\frac{v_i^2}{2\beta_i^{*2}}\right)\times\frac{1}{\sqrt{2\pi e}\beta_k^{*2}}|t_1-t_2|.
	\end{align*}
	We can then take $G=(2\pi)^{-1/2}\underline{c}^{*-2} \exp\{-1/2\}$ according to Condition \ref{eq:gs3}, and $$\omega_\s(v_1,\ldots,v_{(k-1)},v_{(k+1)},\ldots,v_{p_s})= (2\pi)^{-(p_s-1)/2}\left(\prod_{i=1,i\neq k}^{p_s}\beta_i^*\right)^{-1}\exp\left\{-\sum_{i=1,i\neq k}^{p_s}v_i^2(\sqrt{2}\beta_i^{*})^{-2}\right\}.$$ Further, if $\beta^*_r=1$ for $r=1,2,\ldots,p_s$, we have
	$$\int \omega_\s(v_1,\ldots,v_{k-1},v_{k+1},\ldots,v_{p_s})\md v_1\ldots\md v_{k-1}\md v_{k+1}\ldots\md v_{p_s}=1$$ and $$\sum_{l=1,l\neq k}^{p_s}\int |v_{l}| \omega_\s(v_{1},\ldots,v_{k-1},v_{k+1},\ldots,v_{p_s})\md v_{1}\ldots\md v_{k-1}\md v_{k+1}\ldots\md v_{p_s}=(2/\pi)^{(p_s-1)/2}<1.$$
	%

	Condition \ref{eq:last52} requires that the first and second-order derivatives of the density $f_\s(\cdot)$ have a stronger continuous property so that there is enough information around the quasi-true parameter $\sbeta_\s$. When each $x_{\s,r}$ is i.i.d.~$N(0,1)$, we have $|f'_\s(t_1)-f'_\s(t_2)|\leq (2\pi)^{-1/2}|t_1-t_2|$ and $|f''_\s(t_1)-f''_\s(t_2)|\leq \sqrt{2/\pi}|t_1-t_2|$.
	Conditions~\ref{eq:last52} and \ref{eq:last8} are also similarly used in Lemmas .2--.4 of \cite{Ichimura1993Semiparametric} and Condition (ii) of \cite{LiangEstimation}. {Note that we require uniformity across $s$ in this condition because our averaging allows the number of candidate models to diverge}.

	\begin{asu}\label{con:beta2supplement}\
		\begin{itemize}
			\item[] \subasu \label{eq:conbeta2:1}For any $s=1,2,\ldots,S_n$, the objective function $H_{\s,n}(\bbeta_\s)$ defined in \eqref{eq:objective} is twice continuously differentiable;
			\item[] \subasu \label{eq:conbeta2_3}
			There exists a constant $c_0>0$ such that
			\begin{gather}
			\min\left[\infs\lambda_{\min}\left\{\frac{\partial^2 H_{\s,n}(\sbeta_\s)}{\partial \bbeta_\s\partial \bbeta_\s\tt}\right\},\infsj\lambda_{\min}\left\{\frac{\partial^2 H_{\s,n}^{[-j]}(\sbeta_\s)}{\partial \bbeta_\s\partial \bbeta_\s\tt}\right\}\right]\geq c_0>0\notag.
			\end{gather}
		\end{itemize}
	\end{asu}
	
	Condition~\ref{con:beta2supplement} is crucial for the convergence of $\hbeta_\s$ and $\hbeta_\s^{[-j]}$. Condition~\ref{eq:conbeta2:1} is the same as the condition of Lemma~5.4 in \citet{Ichimura1993Semiparametric}, and it requires the smoothness of the objective function. Condition~\ref{eq:conbeta2_3} holds if there exists a local maximum for every $s$ and can be roughly regarded as a minimum-eigenvalue requirement of the ``Fisher information" matrix.
	When $J_n$ and $S_n$ are divergent, we need to restrict the eigenvalues of ``Fisher information" across all blocks and candidate models, and thus $\infs$ and $\min_{1\leq j\leq J_n}$ are needed.
	
	\setcounter{asu}{0}
	
	\subsection{Conditions for Corollary~\ref{thm:high}}
	\label{sec:conditions-col2}
	This section provides the additional conditions required to prove Corollary~\ref{thm:high}.

	\begin{asu}\label{con:high}
		For $s=1,2,\ldots,S_n$ and $r=1,2,\ldots,p_s$, the $r\th$ element of $\hbeta_\s^R$ obtained from \eqref{eq:high_obj}, $\hat{\beta}^R_{\s,r}$, has a limiting value $\beta^{R\ast}_{\s,r}$. Furthermore,
		\begin{align*}
		\sups \left\|\hhbeta_\s-\shbeta_\s\right\|&=O_P(n^{\alpha-1/2}S_n^\gamma),\\
		\sups\supjn \left\|\hhbetaj_\s-\shbeta_\s\right\|&=O_P\left\{(n-M_n)^{\alpha-1/2}S_n^{\gamma}\right\},
		\end{align*}
		where $\alpha\in (0,1/2)$, $\gamma>0$, and $\hhbetaj_\s$ is the CV estimator obtained from the regularized estimation in~\eqref{eq:high_obj}, but excluding observations of the $j{\th}$ block.
	\end{asu}
	
	This condition imposes restrictions on the distances between the quasi-true parameter $\shbeta_\s$ and the two regularized estimators, $\hhbeta_\s$ and $\hhbetaj_\s$, respectively.
	Similar results have been shown by \citet{Radchenko2015High} in a study of regularized estimators for SIMs, and the author focused on the deviation between the estimator and the true parameter as their fitted model coincides with the DGP.
	With $\hhbetaj_\s$ and $\shbeta_\s$ given  above, we can similarly define $\bmu^{R\ast}(\w)$, $\widetilde{\bmu}^R(\w)$ and
	$\widetilde{\bmu}^{R\ast}(\w)$ as the averaging estimators of $\bmu$ using
	$\shbeta_\s$, $\widetilde{\bbeta}_\s^R$ and $\widetilde{\bbeta}^{R\ast}_\s$, respectively. Furthermore, we denote $L_n^{R\ast}(\w)$ as the squared loss of {the regularization-based averaging estimator} and $\xi_n^R=\inf\nolimits_{\w\in \calW}L_n^{R\ast}(\w)$.

	For $s=1,2,\ldots,S_n$, let $\mathcal{C}_s=\left\{r\big|\hat{\beta}^{R}_{\s,r}\neq 0 \text{ or } \beta^{R*}_{\s,r}\neq0\right\}$ and denote the cardinality of $\mathcal{C}_s$ as $q_s$.
	\begin{asu}\label{con:high2}
		\subasu\label{eq:hC61} ${\xi_n^{R}}^{-1}S_n^{\gamma}nq^{1/2}_{\max}(n-M_n)^{\alpha-1/2}=o_P(1)$, where $q_{\max}=\sups q_s$;\\
		\subasu\label{eq:hC62}
		$\xi_n^{R-1}d_n M_n n =o_P(1)$.
	\end{asu}
	
	%
	This condition is a regularization version of Condition~\ref{con:xi}, and provides restrictions on the relative divergent speeds of $S_n$, $\xi_n^{R}$, $q_{\max}$ and $d_n$. Compared with Condition \ref{eq:C61}, Condition \ref{eq:hC61} concerns $q_{\max}$ rather than $p_{\max}$, and it is likely to hold if $q_{\max}$ grows at a slow speed. Note that $q_s$ is related with the ``variable selection consistency", an important property of Lasso-type estimators \citep[see, e.g.,][]{Zou2006The,2010Variable}.

	\begin{asu}\label{con:Hlambda}
		There exists a positive $\rho$ such that, for any $s=1,2,\ldots,S_n$ and any $p_s\times 1$ vector $\mathbf{e}_\s$ consisting of 1 or 0, we have
		\begin{align}
		\lmax\Bigg\{\onen\sumin\left(\mathbf{e}_\s \odot \frac{\partial \hat{g}_\s(\x_{\s,i}\tt\bbeta^{(i)}_\s)}{\partial \bbeta_\s} \right)\left(\frac{\partial \hat{g}_\s(\x_{\s,i}\tt\bbeta^{(i)}_\s)}{\partial \bbeta\tt_\s} \odot\mathbf{e}_\s\tt\right)\Bigg\}=O_P(\|\mathbf{e}_\s\|_1),\label{eq:hlambda}
		\end{align}
		and
		\begin{align}
		\lmax\Bigg\{\onen\sumin\left(\e_\s \odot\frac{\partial \hat{g}_\s^{[-\setB(i)]}(\x_{\s,i}\tt\bbeta^{(i)}_\s)}{\partial \bbeta_\s}\right)\left( \frac{\partial \hat{g}_\s^{[-\setB(i)]}(\x_{\s,i}\tt\bbeta^{(i)}_\s)}{\partial \bbeta\tt_\s}\odot\e_\s\tt\right)\Bigg\}=O_P(\|\e_\s\|_1),\label{eq:hlambda2}
		\end{align}
		{uniformly for any $\bbeta^{(1)}_\s,\bbeta^{(2)}_\s,\ldots,\bbeta^{(n)}_\s\in\mathcal{O}(\sbeta_\s,\rho)$,} where $\odot$ denotes the Hadamard product.
	\end{asu}
	This condition is an extension of Condition \ref{con:lambda}. It degrades to Condition \ref{con:lambda} if $\mathbf{e}_\s\tt=(1,1,\ldots,1)_{1\times p_s}$.

\end{document}